\documentclass[twocolumn, superscriptaddress, secnumarabic, amssymb, amsmath, nofootinbib,tightenlines,nobibnotes, aps, prc]{revtex4-2}

\usepackage{subfigure,dcolumn}
\usepackage[T2A,T1]{fontenc}
\usepackage[russian,english]{babel}
\usepackage{soul, color, xcolor}
\usepackage{listings}
\usepackage{amsmath}
\usepackage{bm}%
\usepackage{prc}
\usepackage[colorlinks,linkcolor=blue,anchorcolor=blue,citecolor=blue]{hyperref}
\usepackage{graphicx}
\usepackage{dcolumn}
\usepackage{color}
\usepackage{booktabs,longtable}
\usepackage{multirow}
\usepackage{CJK}
\usepackage{ltxtable}
\usepackage{rotating}
\usepackage{booktabs}
\usepackage{siunitx}
\makeatletter

\def\@seccntformat#1{\csname the#1\endcsname\quad}
\makeatother

\hypersetup{colorlinks=true, citecolor=blue, linkcolor=blue, urlcolor=blue}
\expandafter\ifx\csname package@font\endcsname\relax\else
 \expandafter\expandafter
 \expandafter\usepackage
 \expandafter\expandafter
 \expandafter{\csname package@font\endcsname}%
\fi

\begin{document}
\title{Evaluation of $^{235,238}$U Fission Product Yields Using Bayesian Neural Networks: Comparison of Baseline and Physics-Informed Models}

\author{Chun-Yuan Qiao}
\affiliation{School of Physics, Centre for Theoretical Physics, Henan Normal University, Xinxiang {\it 453007}, China}
\author{Ya-Xuan Wang}
\affiliation{School of Physics, Centre for Theoretical Physics, Henan Normal University, Xinxiang {\it 453007}, China}
\author{Jun-Chen Pei}
\affiliation{State Key Laboratory of Nuclear Physics and Technology, School of Physics, Peking University, Beijing {\it 100871}, China}
\author{Chun-Wang Ma}\email[Corresponding author.]{E-mail: machunwang@126.com}
\affiliation{School of Physics, Centre for Theoretical Physics, Henan Normal University, Xinxiang {\it 453007}, China}
\affiliation{Institute of Nuclear Science and Technology, Henan Academy of Sciences, Zhengzhou {\it 450046}, China}
\affiliation{Shanghai Research Center for Theoretical Nuclear Physics, NSFC and Fudan University, Shanghai {\it200438}, China}
\author{Yong-Jing Chen}
\affiliation{China Nuclear Data Center, China Institute of Atomic Energy, Beijing {\it 102413}, China}
\author{Jin-Gen Chen}
\affiliation{Shanghai Institute of Applied Physics, Chinese Academy of Sciences, Shanghai {\it 201800}, China
Scopus ID: 57209289876}
\author{Jie Pu}
\affiliation{School of Physics, Centre for Theoretical Physics, Henan Normal University, Xinxiang {\it 453007}, China}
\affiliation{Shanghai Research Center for Theoretical Nuclear Physics, NSFC and Fudan University, Shanghai {\it200438}, China}
\author{Kai-Xuan Cheng}
\affiliation{School of Physics, Centre for Theoretical Physics, Henan Normal University, Xinxiang {\it 453007}, China}
\author{Yu-Ting Wang}
\affiliation{School of Physics, Centre for Theoretical Physics, Henan Normal University, Xinxiang {\it 453007}, China}
\author{Ya-Fei Guo}
\affiliation{School of Physics, Centre for Theoretical Physics, Henan Normal University, Xinxiang {\it 453007}, China}
\author{Xiang Chen}
\affiliation{School of Physics, Centre for Theoretical Physics, Henan Normal University, Xinxiang {\it 453007}, China}

\begin{abstract}
$^{235}$U and $^{238}$U are fundamental materials in thermal and fast neutron breeding studies. Accurate evaluation of their fission product yields is of critical importance for advanced reactor design and nuclear waste management. In this work, a baseline Bayesian neural network model (BNN$_0$) with two hidden layers of 20 neurons each was constructed. An improved model, BNN$_3$, was developed by incorporating additional physics-informed features, namely the odd-even effect, $\beta$-decay energy, and isospin, into the network inputs. Comparative analyses of the general distributions of the fission yields and isotopic chain structures demonstrate that BNN$_3$ exhibits significantly improved reconstruction accuracy and consistency with the target cumulative fission-yield distributions. For 16 representative fission products, the energy-dependent yield predictions of BNN$_3$ show better agreement with both experimental data and evaluated libraries, accompanied by noticeably narrower confidence intervals. These results indicate that the incorporation of relevant physical information improves the model's sensitivity to underlying fission mechanisms and enhances its capability to reproduce the systematic characteristics of cumulative fission-yield distributions. Together, these strategies contribute to more accurate and robust nuclear data modeling, providing a methodological foundation for the evaluation and development of next-generation nuclear data libraries.
\end{abstract}
\maketitle

\section{Introduction}

Nuclear fission, as a fundamental phenomenon in nuclear physics, plays a crucial role in a wide range of applications, including nuclear energy, national defense, and medical isotope production. It is also essential for studies on the synthesis of superheavy nuclei \cite{superheavy1,superheavy2}, the analysis of reactor antineutrino energy spectra \cite{neutrinos_PRD_2011,neutrinos_PRL_2016,neutrinos_PRC_2011}, and the investigation of nucleosynthesis via the {\it r}-process in neutron star mergers \cite{r-process_prl_2013,r-process_apj_2015,r-process_prc_2020}. Accurate and comprehensive fission yield data are critical for understanding the dynamic evolution of the fission process \cite{hill_pr_1953}. Among various fissile and fertile nuclides, $^{235}$U and $^{238}$U have long been the focal points of fission studies due to their significant scientific and practical importance. $^{235}$U is the only naturally occurring fissile isotope with a high fission cross section for thermal neutrons, making it the cornerstone of thermal neutron fission and breeding research \cite{Bohr_pr_1939}. Its fission yield data are essential for isotope production and nuclear waste management, and it also serves as a classic system for exploring macroscopic-microscopic nuclear models, including liquid-drop behavior and shell corrections. In contrast, $^{238}$U is the main fertile nuclide in fast neutron-induced fission and breeding studies \cite{glenn_science_1946,endf}. The evaluation of its fission product yields plays a critical role in the design and optimization of fast neutron nuclear energy systems, as well as in geochronology and studies of natural nuclear reactors.

Currently, the major international nuclear data libraries evaluated (such as JENDL \cite{jendl-url}, ENDF \cite{endf}, CENDL \cite{cendl-Ge2011}, and JEFF \cite{jeff-Plompen2020}) provide comprehensive data for neutron-induced fission yields only at specific energies, that is, 0.0253 eV (thermal neutron), 0.5 MeV, and 14 MeV. However, neutron-induced fission measured data are often sparse, with incomplete isotopic coverage, significant uncertainties,  and notable discrepancies \cite{exfor}. Obtaining complete isotopic yield distributions across a continuous range of incident neutron energies remains experimentally challenging. Phenomenological models of nuclear fission (such as the Brosa \cite{brosa} and GEF \cite{gef} approaches) rely heavily on experimental data for parameter calibration and validation \cite{schunck_2016}. When such data are limited or absent, the underlying assumptions and approximations of these models become weakly constrained, resulting in increased parameter uncertainties and diminished predictive reliability. 
 
With the development of artificial intelligence, machine learning (ML) techniques have found extensive and diverse applications in nuclear physics research \cite{zhang_jpg_2017,Boehnlein_rmp_2022,pedro_ANE_2021}. These applications encompass a wide range of studies, including nuclear mass prediction for advancing nuclear theory \cite{niu_plb_2018,YANG_PLB_2021,wuxh_PRC_2020,MING_NST_2022}, astrophysical modeling \cite{Utama_PRC_2016,Utama_PRC_2017,niu_prc_2022}, and methodological comparisons \cite{niu_prc_2019,liu_prc_2021}; determination of nuclear charge radii \cite{wu_prc_2020,Akkoyun_jpg_2013,radii-Utama_2016,radii_PRCMA_2020,radii_NST_2023}; estimation of half-lives of $\alpha$-decay and $\beta$-decay \cite{alpha_PRC_2022,beta_PRC_2019,beta_PRCLi_2025}; the inference of spin parities in the ground-state \cite{gradespin-Liu2024}; and prediction of fission product yields \cite{WAGN_PRL_2019,lovell_epj_2019,SONG_NST_2023,SONG_PRC_2023,wyx}. Furthermore, in the realm of nuclear reactions, ML has been effectively employed to model cross sections in various mechanisms, such as proton-induced spallation \cite{pengdan_jpg_2022,pengdan_scisin_2022,ma_cpc_2020_12,ma_cpc_2020_1}, projectile fragmentation \cite{ma_cpc_2022,ma_prc_2023,wei_prc_2025,weixb_nst_2022}, and photonuclear reactions \cite{li_nst_2022,sun_nst_2025}.
ML algorithms can uncover hidden patterns in complex datasets and extract the underlying physical correlations.
Among them, the Bayesian neural network (BNN) has demonstrated strong potential to evaluate fission product yields \cite{WAGN_PRL_2019} and investigate their dependence on neutron energy \cite{qiao_sci_2022}. Unlike traditional neural networks, BNN treats the model parameters as probability distributions rather than fixed values. This allows not only yield predictions but also quantifying associated uncertainties \cite{phillips_jpg_2021}. By incorporating prior knowledge and posterior inference, BNN imposes meaningful constraints on model parameters and effectively mitigates overfitting risks \cite{kegzlar_jpg_2020}. Moreover, BNN naturally integrates experimental uncertainties and missing data during network training, thus improving both the robustness and predictive accuracy of the model \cite{wang_prc_2022,yijy_scisin_2022}.

In this study, a Bayesian neural network (BNN) architecture consisting of two hidden layers with 20 neurons each and employing the \textit{tanh} activation function \cite{qiao_sci_2022, wang_prc_2021} was adopted to evaluate the fission product yields of $^{235,238}$U. The initial training set inputs include the proton number, mass number of the fission fragments, and the excitation energy of the compound nucleus, which forms the baseline model BNN$_0$. To further enhance the predictive capability of the network~\cite{jia_2021, ml_2021, hillips_jpg_2021, ml_prc_2021}, additional physical quantities such as the odd-even effect, $\beta$-decay energy, and isospin were incorporated. After normalization preprocessing, an improved model, BNN$_3$, was constructed. The BNN$_3$ model demonstrated significantly better predictive and generalization performance than BNN$_0$, particularly in reproducing the overall shape of the fission fragment yield distributions and the isotope chain yields. Furthermore, the energy dependence of the cumulative yields for 16 fission products was presented. Compared to BNN$_0$, BNN$_3$ exhibited better consistency with both evaluated and experimental data, while producing sharper predictive intervals and maintaining confidence interval coverage close to the nominal level.

The remainder of this paper is organized as follows. Section~\ref{BNN model} provides a comprehensive overview of the theoretical framework of the BNN approach, along with the composition and preprocessing of the training dataset used for model construction. Section~\ref{results} provides a detailed analysis of the results, with Section~\ref{results-U235} focusing on $^{235}$U and Section~\ref{results-U238} on $^{238}$U. Finally, Section~\ref{summary} summarizes the key findings and methodological innovations of this work.

\section{BNN model}\label{BNN model}
A Bayesian neural network (BNN) is adopted in this study to predict fission product yields by combining Bayesian inference with the structure of traditional neural networks \cite{neal2012bayesian}. Unlike conventional neural networks, where model parameters are fixed after training and produce a single deterministic output, BNN represents model parameters as probability distributions. This probabilistic formulation enables BNN to provide both predictions and reliable estimates of associated uncertainties.

\begin{figure}[htbp]\centering    \includegraphics[width=0.48\textwidth]{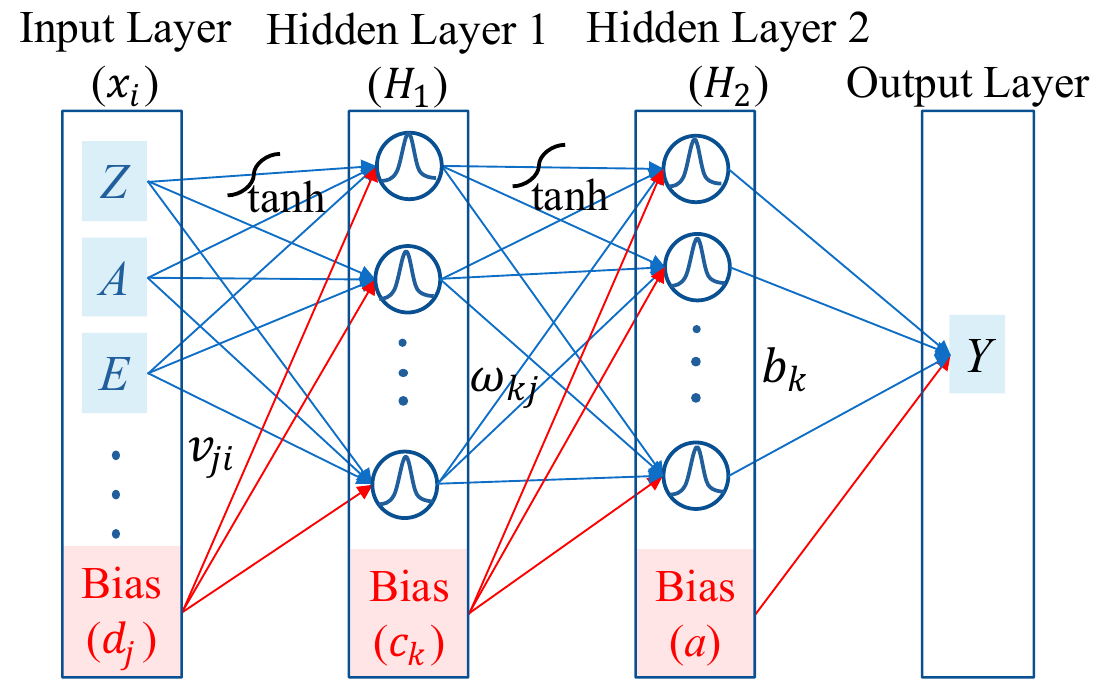} 
    \caption{(Color online) {Schematic of the two-hidden-layer Bayesian neural network for fission yield evaluation. The input layer includes the proton numbers and mass numbers of the fission fragments, as well as compound nucleus excitation energy ($e_\mathrm{n} + S_\mathrm{n}$). The network consists of two hidden layers with $H_1 = 20$ and $H_2 = 20$ neurons, respectively, both using $\tanh$ activation functions. The output layer provides the fission yield. All connection weights ($v_{ji}$, $w_{kj}$, $b_k$) and biases ($d_j$, $c_k$, $a$) are treated as probability distributions within the Bayesian framework.}}
    \label{fig1-net}
\end{figure}
\begin{figure}[htbp]\centering    \includegraphics[width=0.48\textwidth]{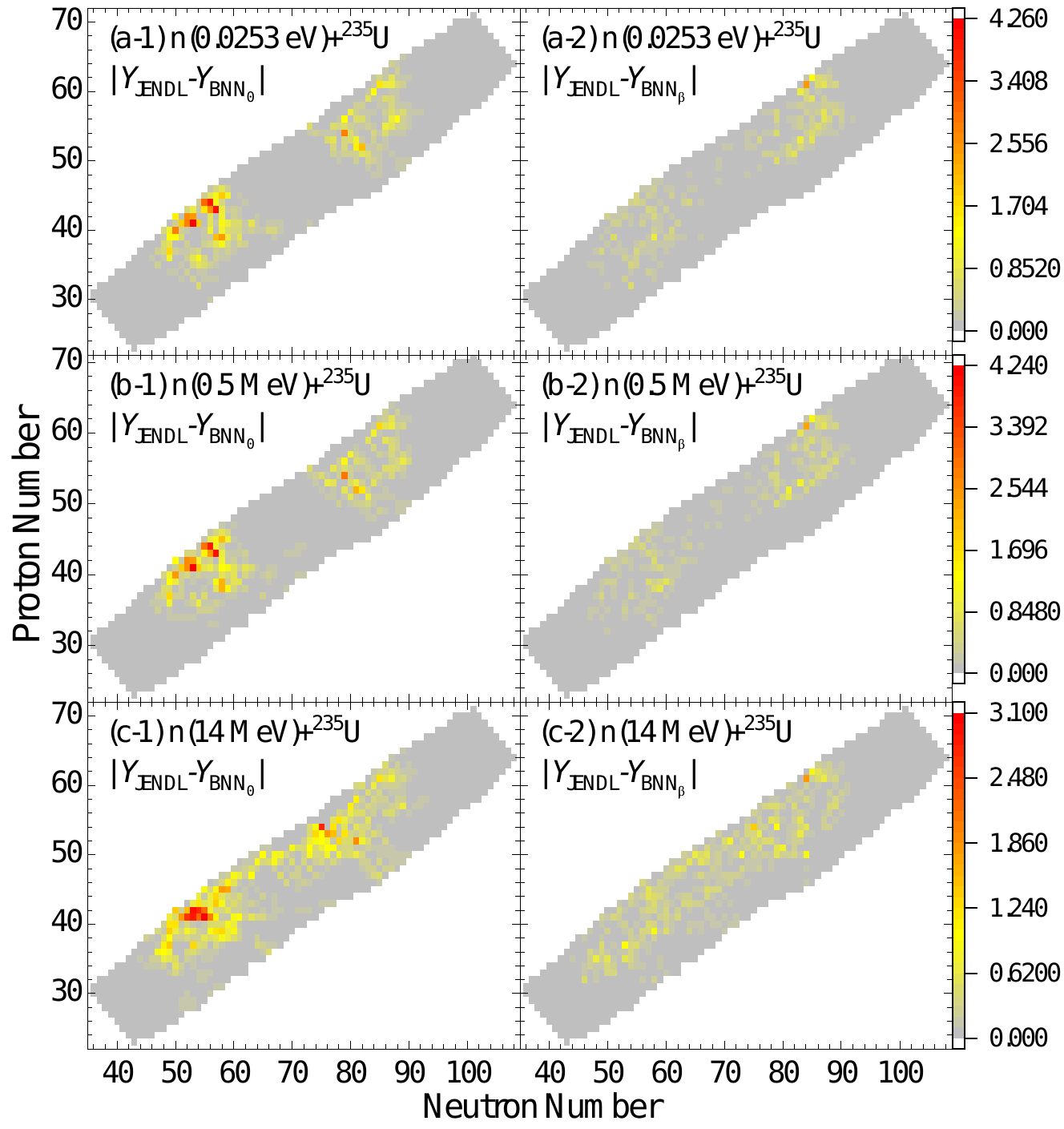} 
    \caption{(Color online) {Distributions of absolute residuals between the JENDL evaluated cumulative fission yields and Bayesian neural network predictions for neutron-induced fission of $^{235}$U. The quantities displayed are $|Y_{\mathrm{JENDL}}-Y_{\mathrm{BNN_0}}|$ in panels (a-1), (b-1), and (c-1), and $|Y_{\mathrm{JENDL}}-Y_{\mathrm{BNN_{\beta}}}|$ in panels (a-2), (b-2), and (c-2). $Y_{\mathrm{BNN_0}}$ denotes the prediction of the baseline Bayesian neural network trained with the original input features, whereas $Y_{\mathrm{BNN_{\beta}}}$ denotes the prediction of the network incorporating the $\beta$-decay energy as an additional input feature. Panels (a), (b), and (c) correspond to incident neutron energies of 0.0253 eV, 0.5 MeV, and 14 MeV, respectively.}}
    \label{beta1}
\end{figure}
\begin{figure*}[htbp]
    \centering
    \includegraphics[width=0.98\textwidth]{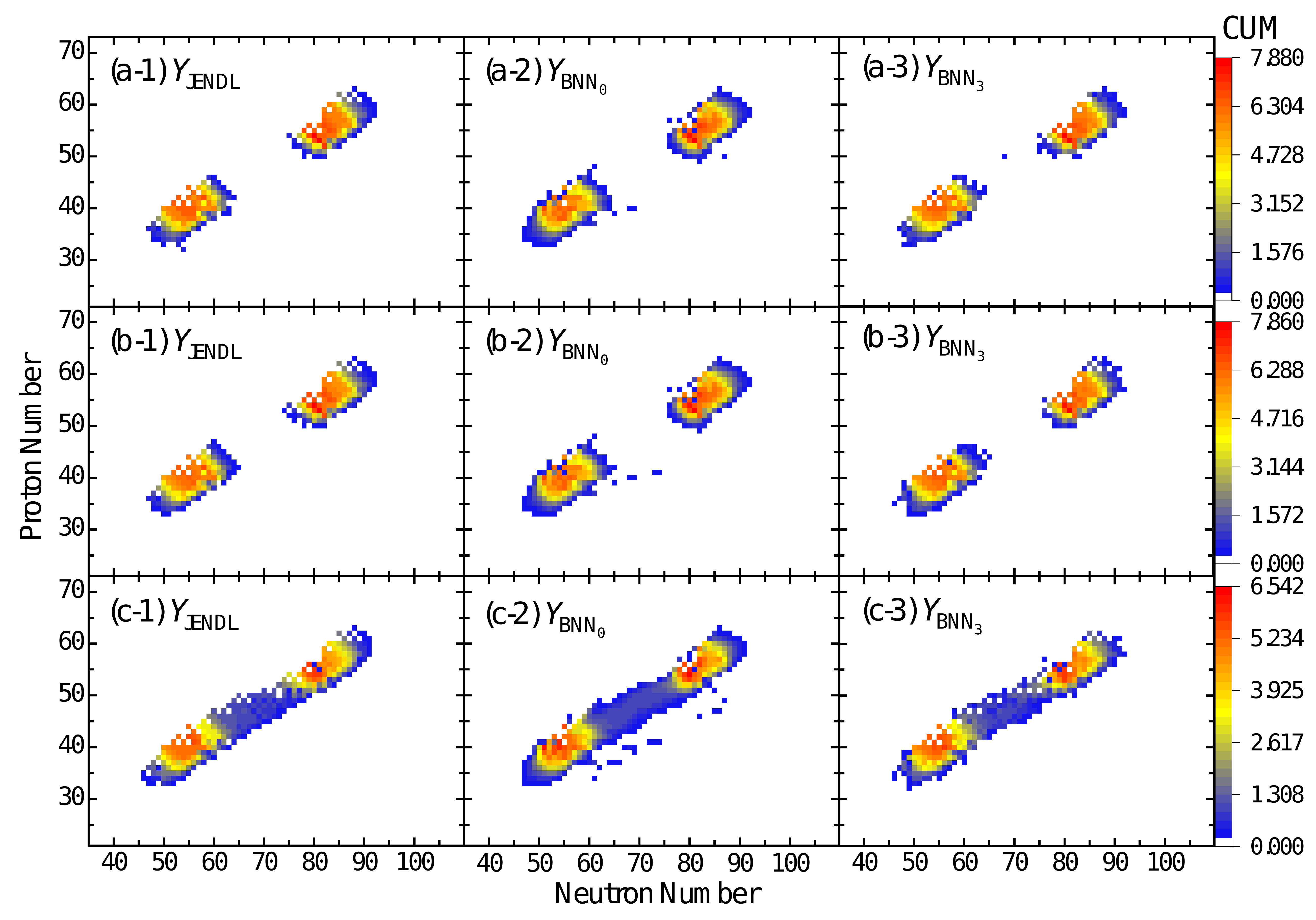} 
    \caption{(Color online) Cumulative fission product yield distributions of neutron-induced fission of $^{235}$U based on different data sources. $Y_{\mathrm{JENDL}}$ denotes evaluated yields from the JENDL library. $Y_{\mathrm{BNN}_0}$ represents predictions from a Bayesian neural network trained with the original dataset, while $Y_{\mathrm{BNN}_3}$ corresponds to predictions from a network incorporating the additional physical information, including the odd-even effect, $\beta$-decay energy, and isospin. Panels (a-1) to (a-3) correspond to fission induced by thermal neutrons (0.0253~eV), (b-1) to (b-3) to fission induced by 0.5~MeV neutrons, and (c-1) to (c-3) to fission induced by 14~MeV neutrons.}
    \label{fig2-U235-heatmap}
\end{figure*}

\begin{figure}[htbp]
    \centering
    \includegraphics[width=0.48\textwidth]{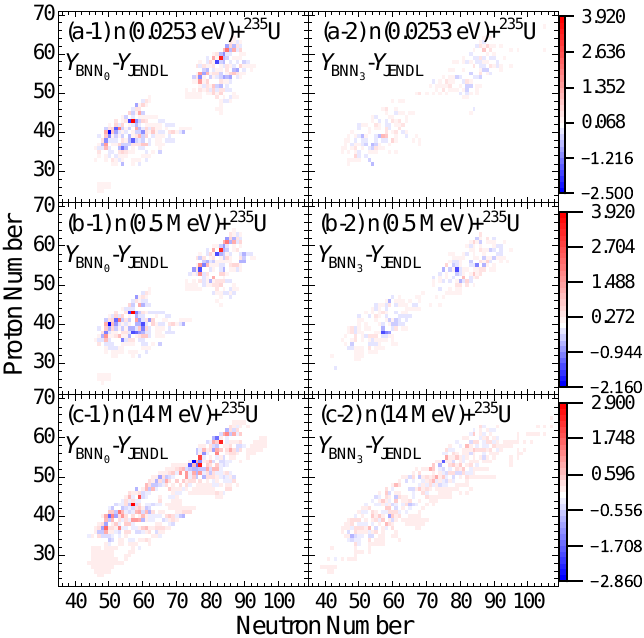} 
    \caption{(Color online) Residual distributions between the BNN predictions and the JENDL evaluated cumulative fission yields for neutron-induced fission of \( ^{235}\mathrm{U} \) at three incident neutron energies: thermal, 0.5 MeV, and 14 MeV.}
    \label{fig-235-error}
\end{figure}
\begin{figure*}[htbp]
    \centering
    \includegraphics[width=0.98\textwidth]{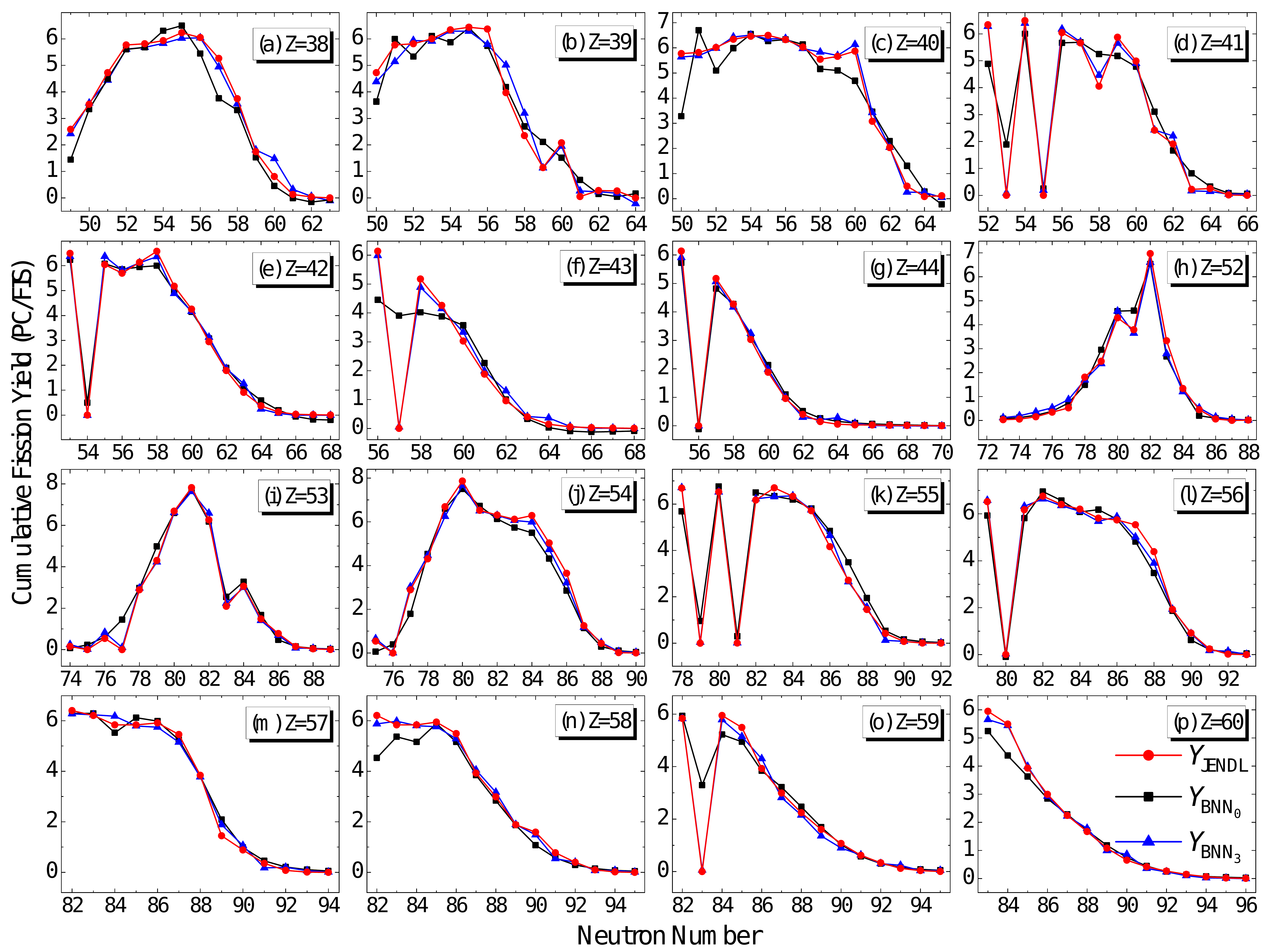} 
    \caption{(Color online) Cumulative fission yield distributions of isotopic chains with proton numbers $Z = 38$-44 and $Z = 52$-60 from thermal neutron-induced fission of \( ^{235}\mathrm{U} \). The red circles (\( Y_\mathrm{JENDL} \)) denote evaluated data from the JENDL database, the black squares (\( Y_\mathrm{BNN_0} \)) represent predictions by a Bayesian neural network trained on the original dataset, and the blue triangles (\( Y_\mathrm{BNN_3} \)) show predictions from a network trained with additional inputs including odd-even effect, $\beta$-decay energy, and isospin.}
    \label{fig3-U235-0.0253eV}
\end{figure*}
\begin{figure*}[htbp]
    \centering
    \includegraphics[width=0.98\textwidth]{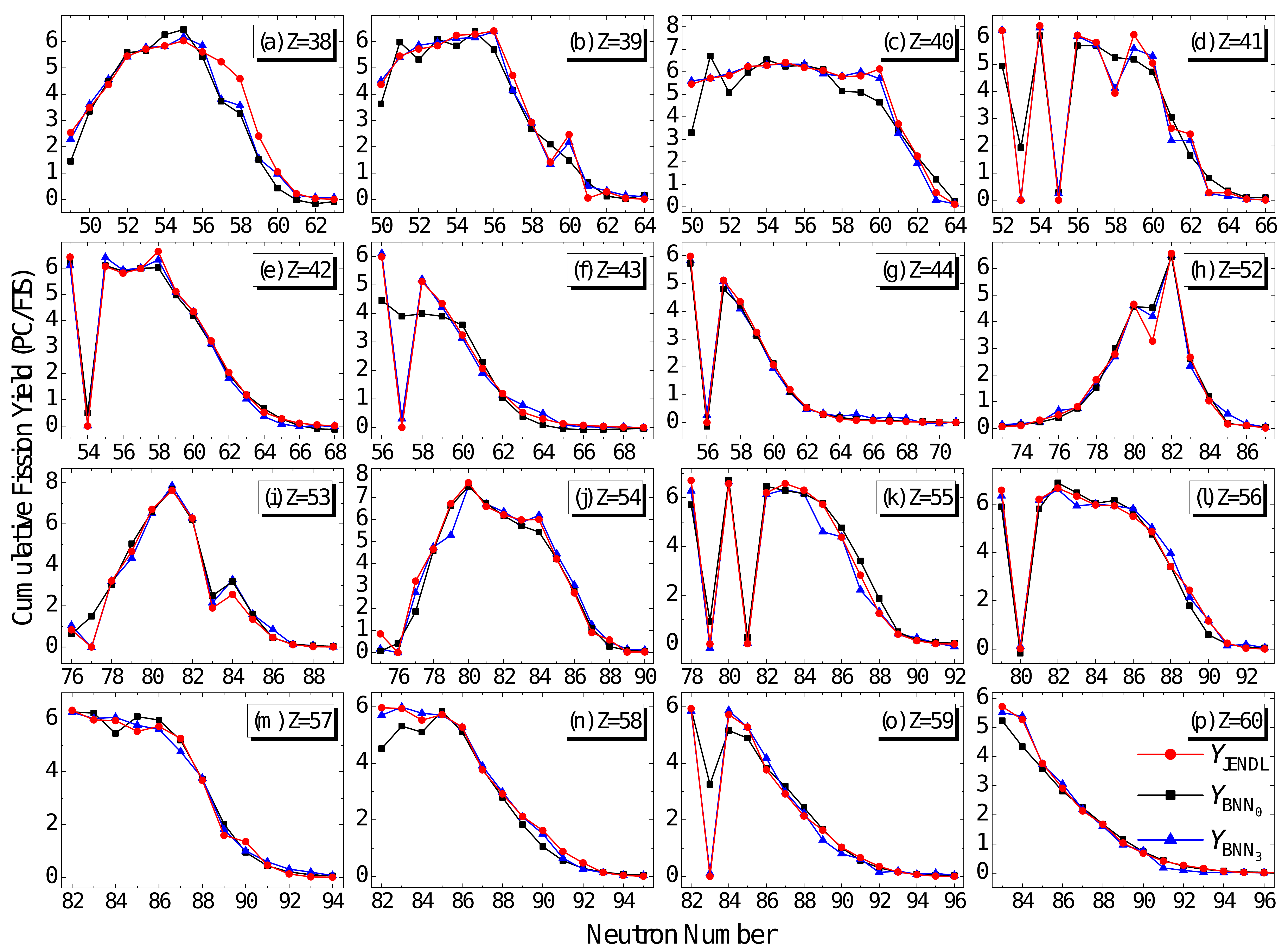} 
    \caption{(Color online) Cumulative fission yield distributions of isotopic chains with proton numbers $Z = 38$-44 and $Z = 52$-60 from 0.5~MeV neutron-induced fission of \( ^{235}\mathrm{U} \). The red circles (\( Y_\mathrm{JENDL} \)) denote evaluated data from the JENDL database, the black squares (\( Y_\mathrm{BNN_0} \)) represent predictions by a Bayesian neural network trained on the original dataset, and the blue triangles (\( Y_\mathrm{BNN_3} \)) show predictions from a network trained with additional inputs including odd-even effect, $\beta$-decay energy, and isospin.}
    \label{fig4-U235-0.5MeV}
\end{figure*}
\begin{figure*}[htbp]
    \centering
    \includegraphics[width=0.98\textwidth]{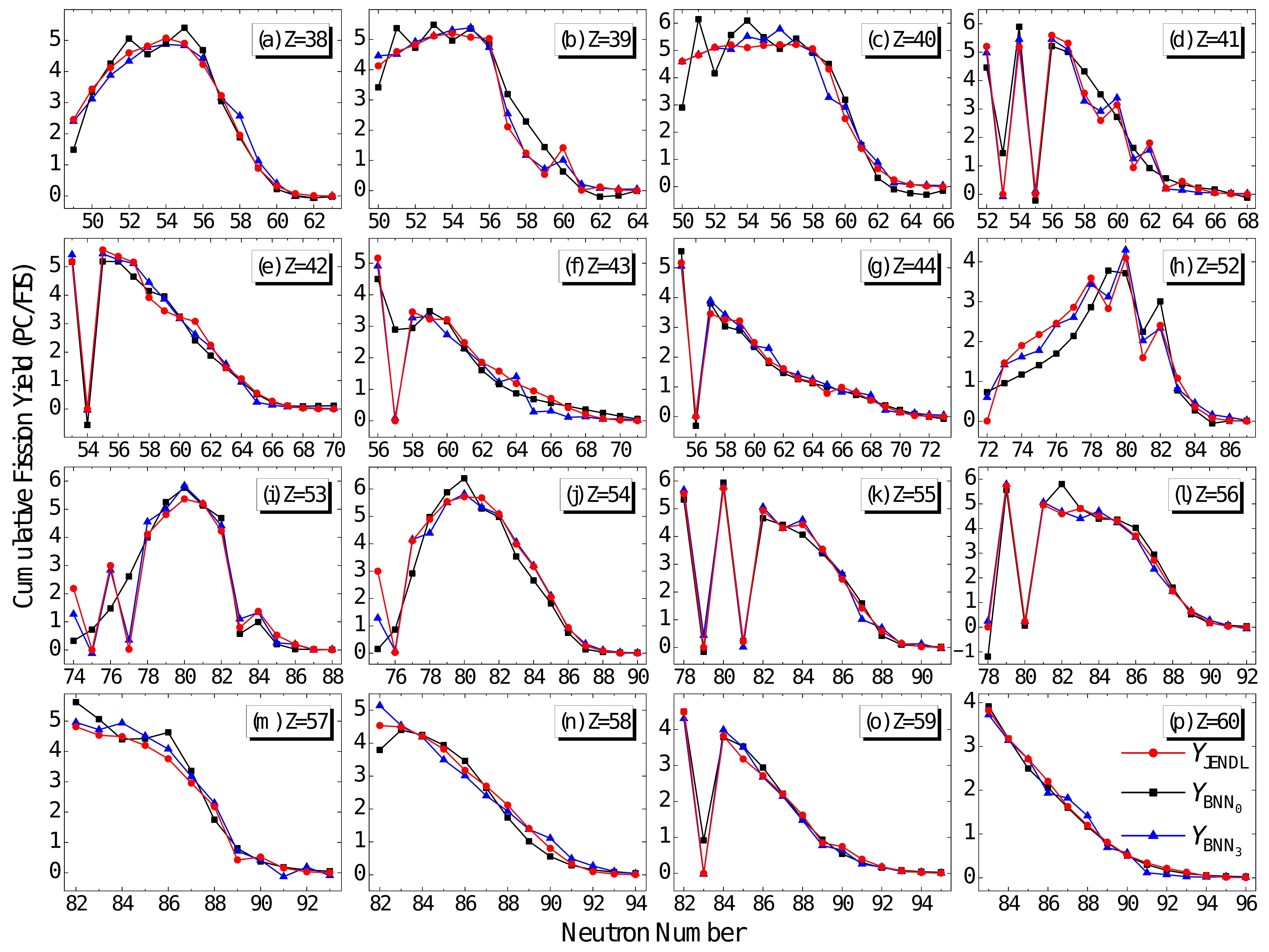} 
    \caption{(Color online) Cumulative fission yield distributions of isotopic chains with proton numbers $Z = 38$-44 and $Z = 52$-60 from 14~MeV neutron-induced fission of \( ^{235}\mathrm{U} \). The red circles (\( Y_\mathrm{JENDL} \)) denote evaluated data from the JENDL database, the black squares (\( Y_\mathrm{BNN_0} \)) represent predictions by a Bayesian neural network trained on the original dataset, and the blue triangles (\( Y_\mathrm{BNN_3} \)) show predictions from a network trained with additional inputs including odd-even effect, $\beta$-decay energy, and isospin.}
    \label{fig5-U235-14MeV}
\end{figure*}
\begin{figure*}[htbp]
    \centering
    \includegraphics[width=0.98\textwidth]{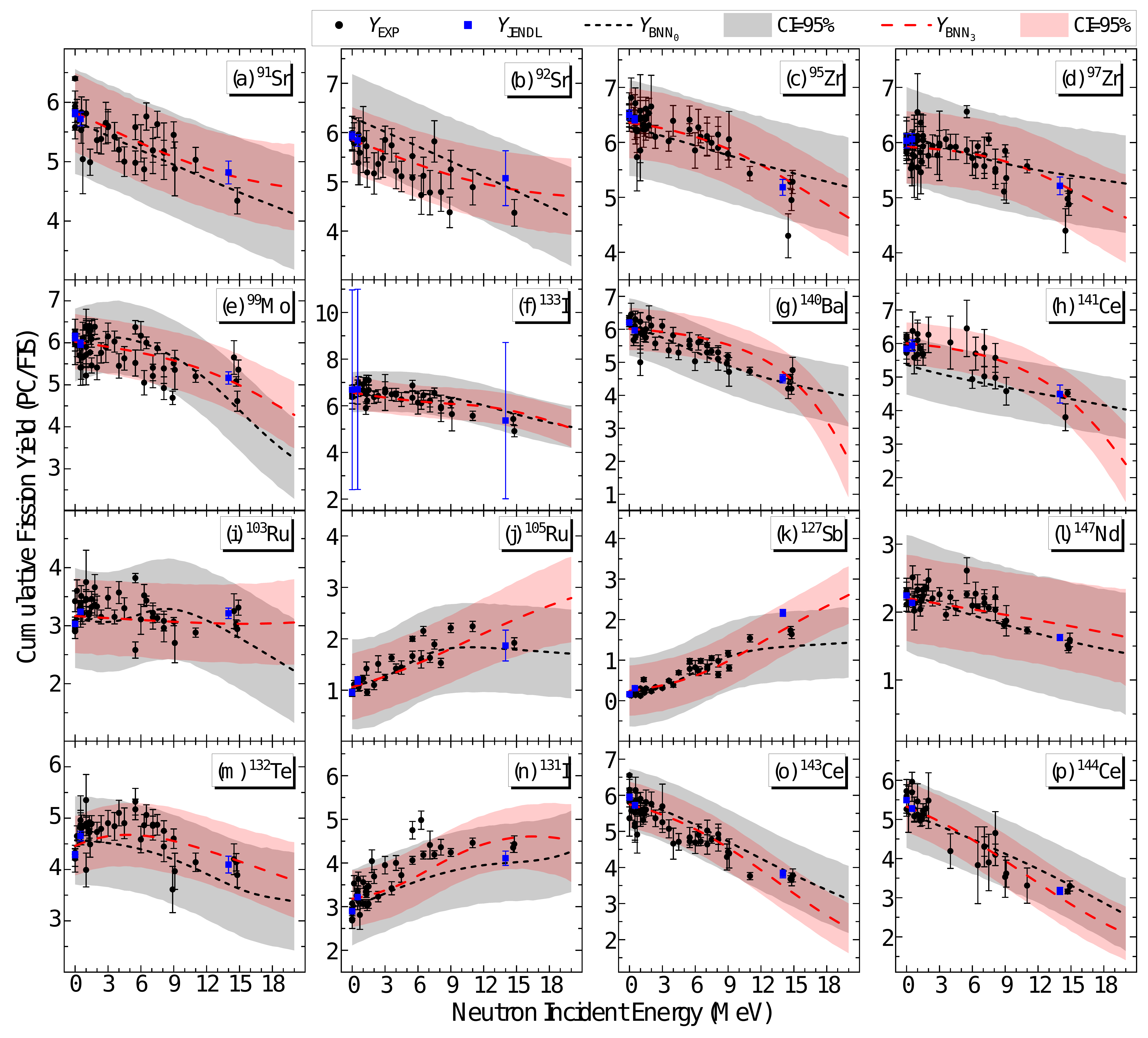} 
    \caption{(Color online) Energy dependence of cumulative fission yields for 16 fission products from neutron-induced fission of \( ^{235}\mathrm{U} \) in the incident neutron energy range of 0-20~MeV. Black circles (\( Y_\mathrm{EXP} \)) represent experimental data from the EXFOR database, blue squares (\( Y_\mathrm{JENDL} \)) denote evaluated yields from JENDL, black short dashed lines (\( Y_\mathrm{BNN_0} \)) show predictions from a Bayesian neural network trained on the original dataset, and red dashed lines (\( Y_\mathrm{BNN_3} \)) correspond to predictions from a network trained with additional inputs including the odd-even effect, $\beta$-decay energy, and isospin. The shaded bands represent the 95\% confidence intervals of the BNN predictions.}
    \label{fig6-U235-yieldenergy}
\end{figure*}
\begin{figure*}[htbp]
    \centering
    \includegraphics[width=0.98\textwidth]{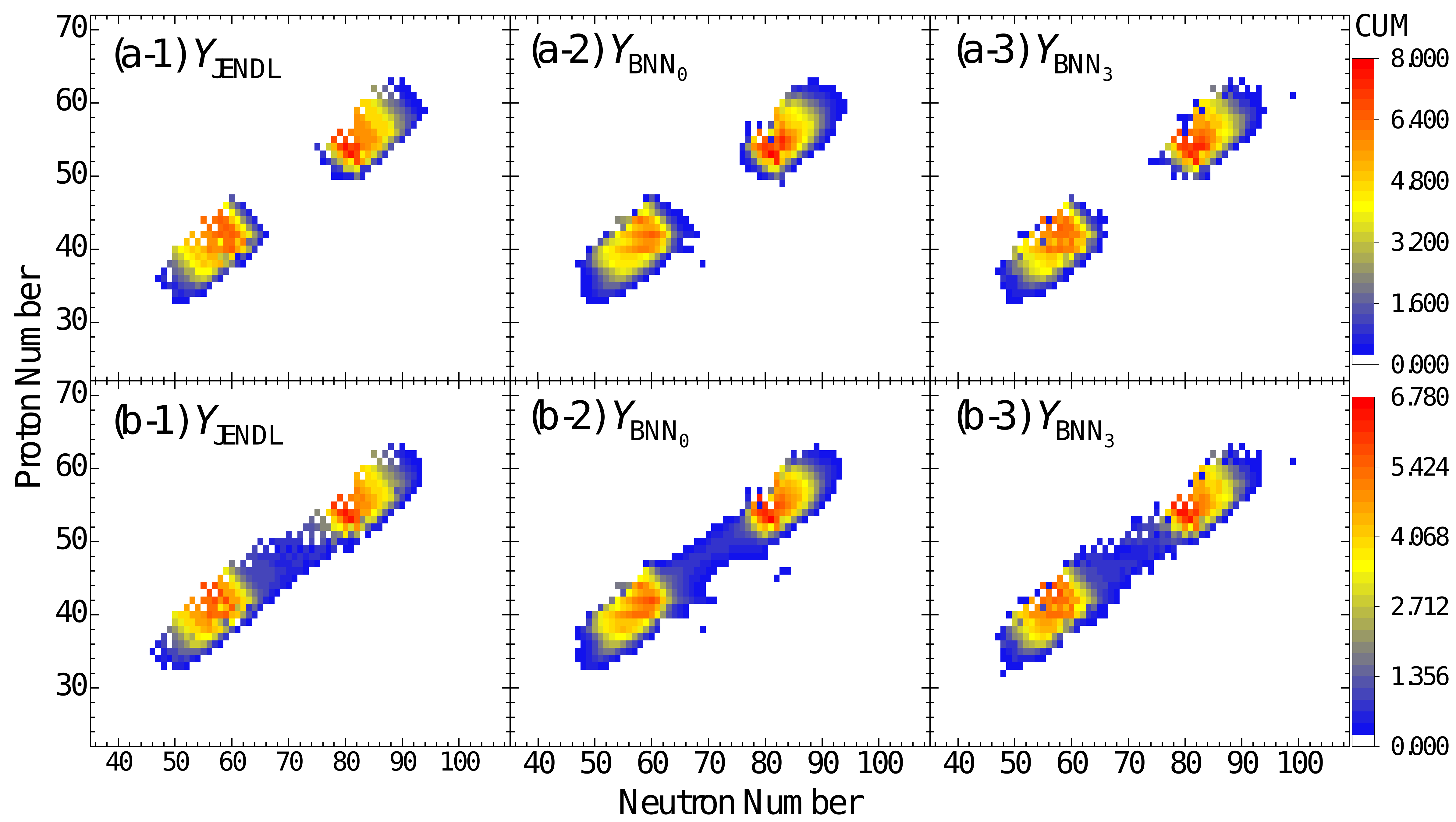} 
    \caption{(Color online) Cumulative fission product yield distributions of neutron-induced fission of $^{238}$U from different data sources. $Y_{\mathrm{JENDL}}$ refers to the evaluated data from the JENDL library. $Y_{\mathrm{BNN}_0}$ is predicted by a Bayesian neural network trained on the original dataset, while $Y_{\mathrm{BNN}_3}$ includes additional physical features: the odd-even effect, $\beta$-decay energy, and isospin. Panels (a-1) to (a-3) show results for 0.5~MeV neutrons, and (b-1) to (b-3) for 14~MeV neutrons.}
    \label{fig7-U238-heatmap}
\end{figure*}
\begin{figure}[htbp]
    \centering
    \includegraphics[width=0.48\textwidth]{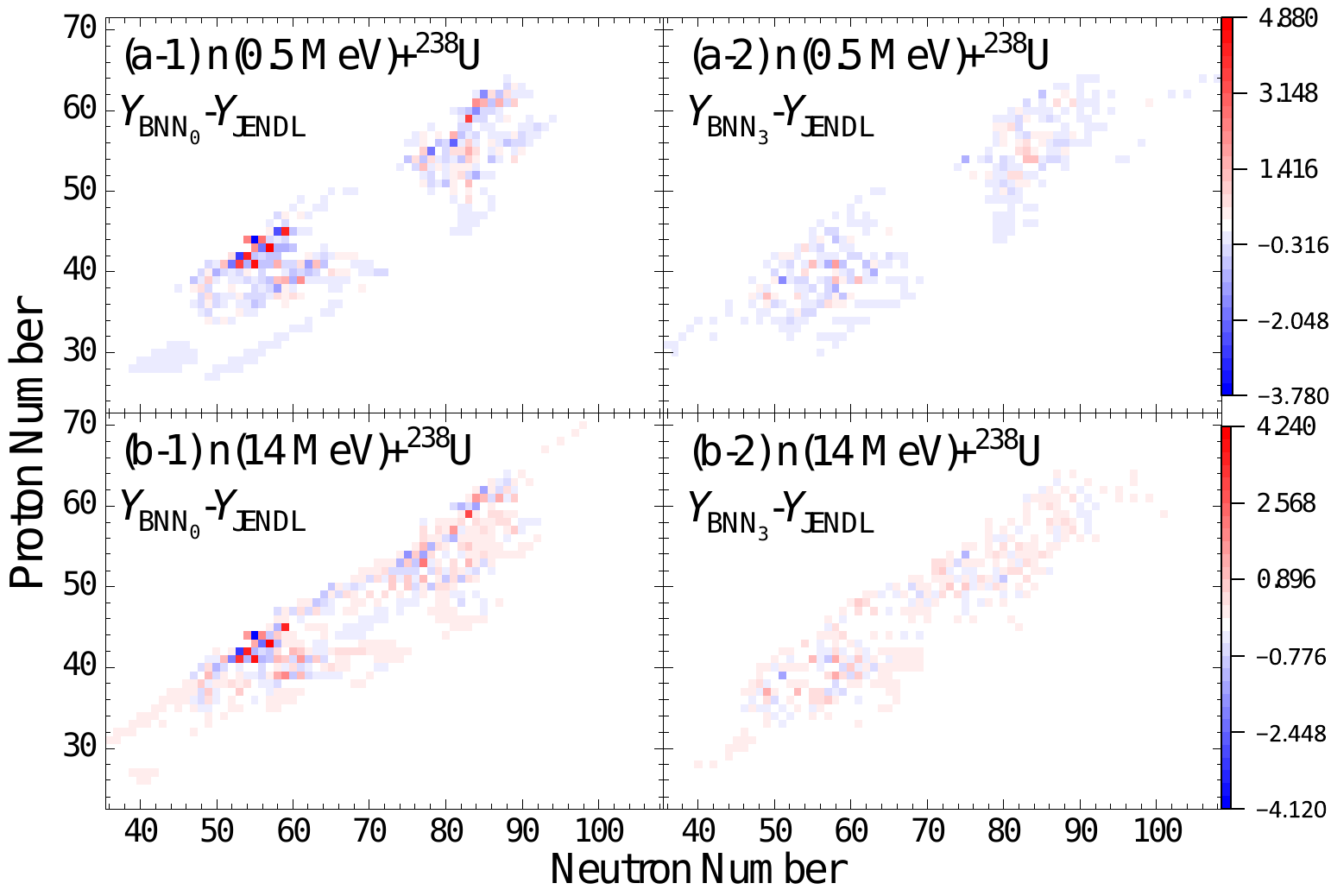} 
    \caption{(Color online) Residual distributions between the BNN predictions and the JENDL evaluated cumulative fission yields for neutron-induced fission of \( ^{238}\mathrm{U} \) at incident neutron energies of 0.5 MeV and 14 MeV.}
    \label{fig-238-error}
\end{figure}
\begin{figure*}[htbp]
    \centering
    \includegraphics[width=0.98\textwidth]{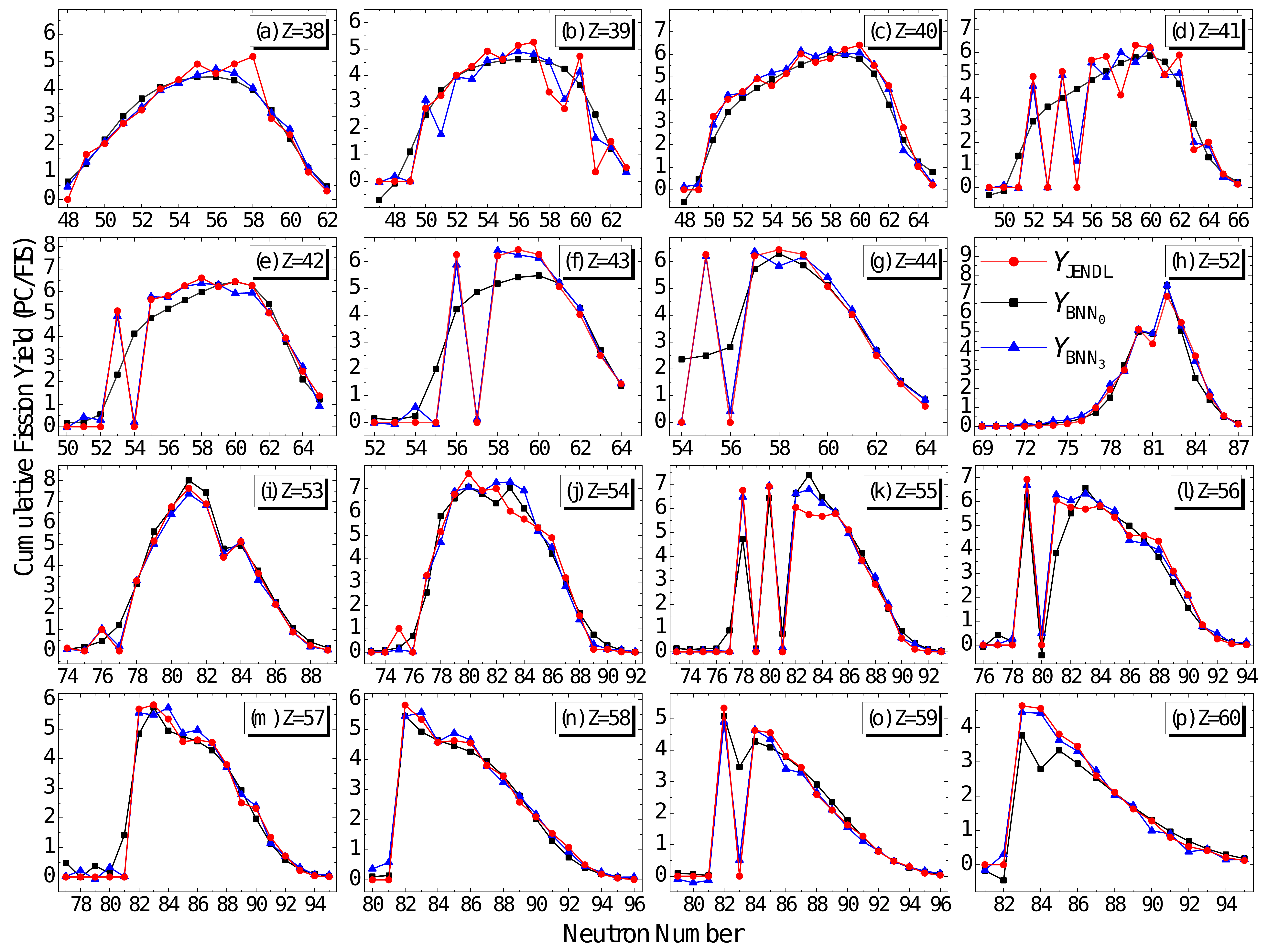} 
    \caption{(Color online) Cumulative fission yield distributions of isotopic chains with proton numbers $Z = 38$-44 and $Z = 52$-60 from 0.5~MeV neutron-induced fission of \( ^{238}\mathrm{U} \). The red circles (\( Y_\mathrm{JENDL} \)) denote evaluated data from the JENDL database, the black squares (\( Y_\mathrm{BNN_0} \)) represent predictions by a Bayesian neural network trained on the original dataset, and the blue triangles (\( Y_\mathrm{BNN_3} \)) show predictions from a network trained with additional inputs including odd-even effect, $\beta$-decay energy, and isospin.}
    \label{fig8-U238-0.5MeV}
\end{figure*}
\begin{figure*}[htbp]
    \centering
    \includegraphics[width=0.98\textwidth]{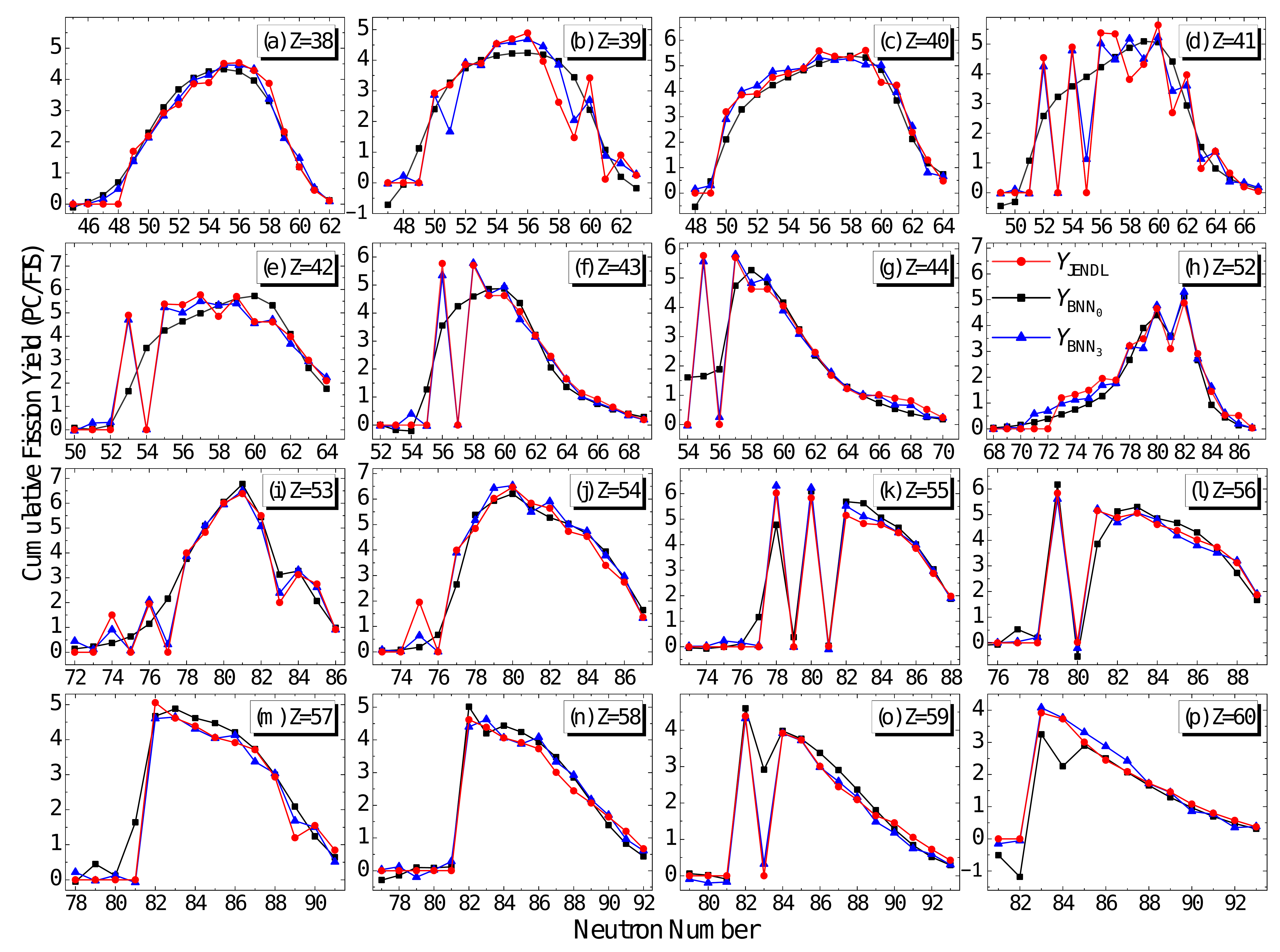} 
    \caption{(Color online) Cumulative fission yield distributions of isotopic chains with proton numbers $Z = 38$-44 and $Z = 52$-60 from 14~MeV neutron-induced fission of \( ^{238}\mathrm{U} \). The red circles (\( Y_\mathrm{JENDL} \)) denote evaluated data from the JENDL database, the black squares (\( Y_\mathrm{BNN_0} \)) represent predictions by a Bayesian neural network trained on the original dataset, and the blue triangles (\( Y_\mathrm{BNN_3} \)) show predictions from a network trained with additional inputs including odd-even effect, $\beta$-decay energy, and isospin.}
    \label{fig9-U238-14MeV}
\end{figure*}
\begin{figure*}[htbp]
    \centering
    \includegraphics[width=0.98\textwidth]{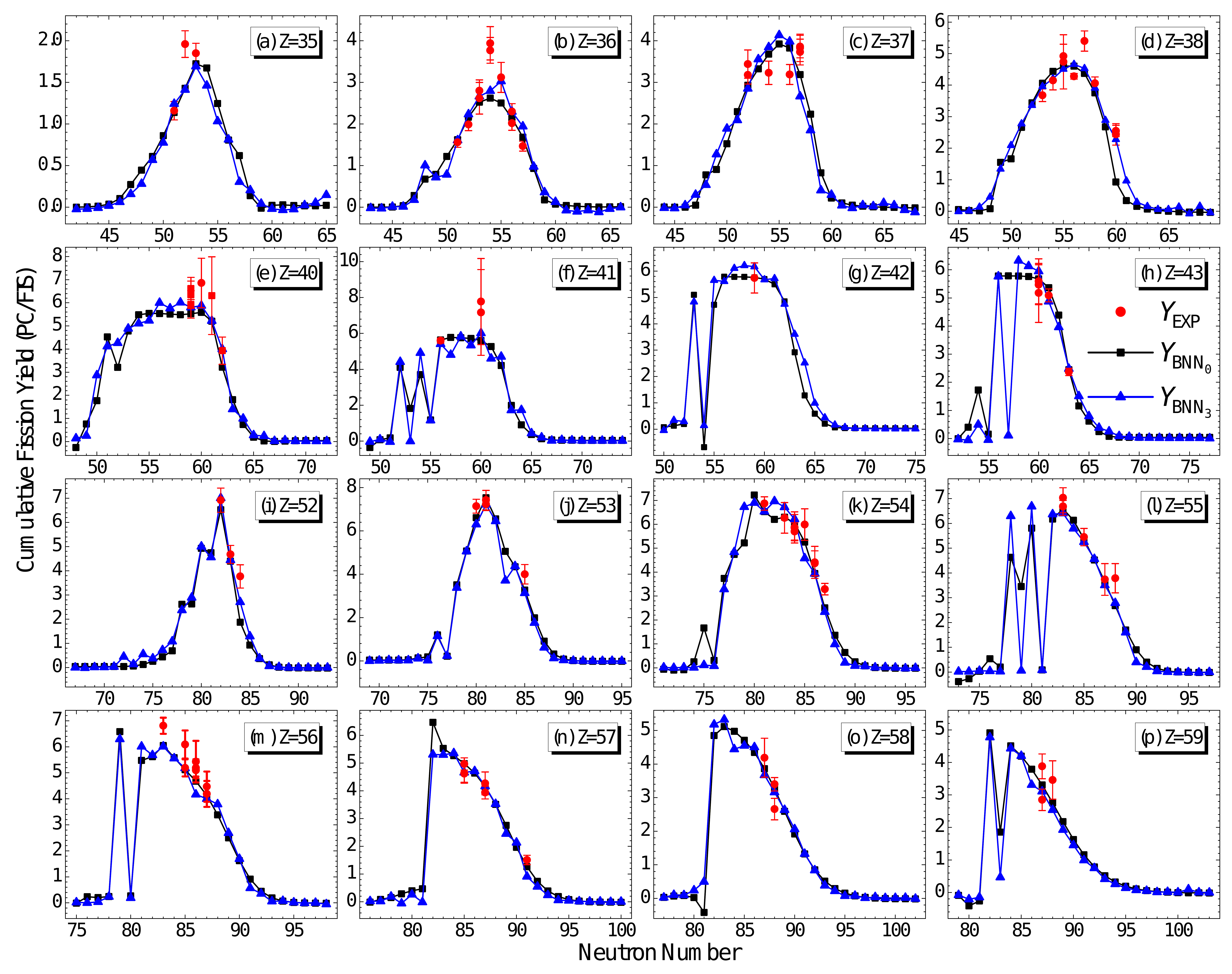} 
    \caption{(Color online) Cumulative fission yield distributions of isotopic chains with proton numbers $Z = 35$-38, $Z = 40$-43, and $Z = 52$-59 from 4.6~MeV neutron-induced fission of \( ^{238}\mathrm{U} \). The red circles (\( Y_\mathrm{EXP}\)) represent experimental data from Ref.~\cite{exp238-4.6}, the black squares (\( Y_\mathrm{BNN_0} \)) represent predictions by a Bayesian neural network trained on the original dataset, and the blue triangles (\( Y_\mathrm{BNN_3} \)) show predictions from a network trained with additional inputs including odd-even effect, $\beta$-decay energy, and isospin.} 
    \label{fig10-U238-4.6MeV}
\end{figure*}
\begin{figure*}[htbp]
    \centering
    \includegraphics[width=0.98\textwidth]{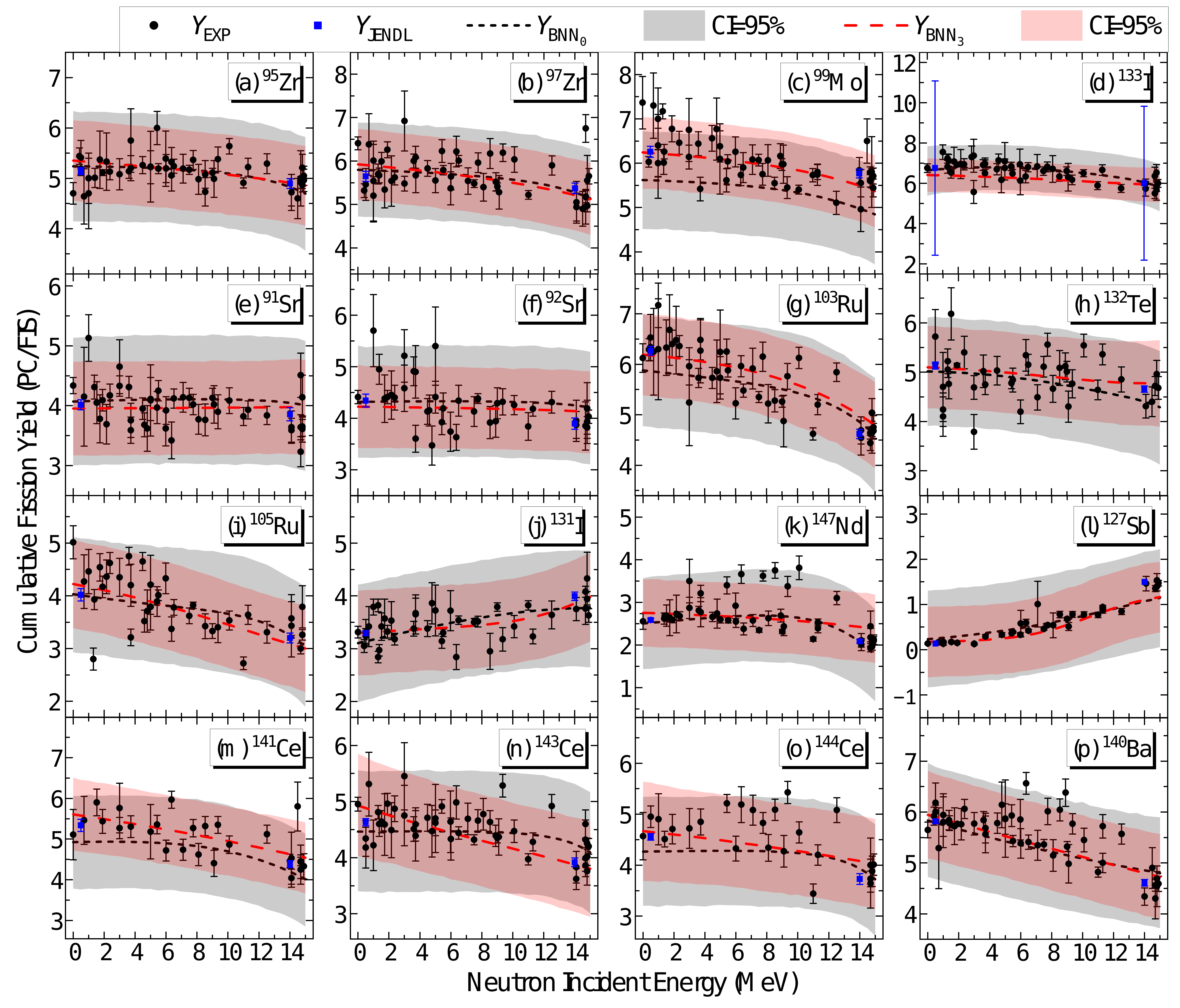} 
    \caption{(Color online) Energy dependence of cumulative fission yields for 16 fission products from neutron-induced fission of \( ^{238}\mathrm{U} \) in the incident neutron energy range of 0-15~MeV. Black circles (\( Y_\mathrm{EXP} \)) represent experimental data from the EXFOR database, blue squares (\( Y_\mathrm{JENDL} \)) denote evaluated yields from JENDL, black short dashed lines (\( Y_\mathrm{BNN_0} \)) show predictions from a Bayesian neural network trained on the original dataset, and red dashed lines (\( Y_\mathrm{BNN_3} \)) correspond to predictions from a network trained with additional inputs including the odd-even effect, $\beta$-decay energy, and isospin. The shaded bands represent the 95\% confidence intervals of the BNN predictions.}
    \label{fig11-U238-yieldenergy}
\end{figure*}

The core of BNN lies in the posterior probability distribution of the model parameters,
\begin{equation}
\mathrm{p}(\theta | x,t) = \frac{\mathrm{p}(x,t|\theta) \mathrm{p}(\theta)}{\int \mathrm{p}(x,t| \theta) \mathrm{p}(\theta) \mathrm{d}\theta},
\end{equation}
where \(\mathrm{p}(x,t|\theta)\) is the likelihood function that describes the probability of observed data given the model parameters \(\theta\), and \(\mathrm{p}(\theta)\) is the prior distribution that represents previous beliefs about these parameters. $\int \mathrm{p}(x,t|\theta)\mathrm{p}(\theta)\mathrm{d}\theta$ serves as the normalization constant to ensure that the posterior is a valid probability distribution. As training data accumulates, the influence of the prior diminishes in favor of the likelihood.

The likelihood function for an individual sample is given by,
\begin{equation}
\begingroup
\begin{aligned}
p(t_i\mid x_i,\theta)
=
\frac{1}{\sqrt{2\pi(\delta_i^2+(\Delta t_i)^2)}}
\exp
\left[
-\frac{(t_i-f(x_i,\theta))^2}
{2(\delta_i^2+(\Delta t_i)^2)}
\right].
\end{aligned}
\endgroup
\end{equation}

The likelihood function for the complete dataset can be expressed as,
\begin{equation}
\begingroup
\begin{aligned}
p(t\mid x,\theta)
=
\prod_{i=1}^{N}
p(t_i\mid x_i,\theta).
\end{aligned}
\endgroup
\end{equation}

The loss function is defined as,
\begin{equation}
\begingroup
\begin{aligned}
\chi^2(\theta) = \sum_{i=1}^{N} \frac{\left({t_i-f(x_i,\theta)}\right)^2}{\delta_i^2+(\Delta t_i)^2}, 
\end{aligned}
\endgroup
\end{equation}
where \(t_i\) is the target value, \(f(x_i, \theta)\) is the prediction of the network for input \(x_i\), and $\delta _i$ represents an adaptive noise term that is uniformly initialized but dynamically optimized during training iterations, while $\Delta t_{i}$ denotes the uncertainty of each fission yield. This mechanism prevents the network from over-relying on low-precision data while still incorporating all available information for robust fission yield evaluation.

Figure~\ref{fig1-net} shows the schematic structure of the two-hidden-layer Bayesian neural network used for fission yield evaluation. The network adopts the tanh activation function and consists of two hidden layers, each containing 20 neurons. In Bayesian neural networks, a feed-forward neural network is used, 
\begin{equation}
\begingroup
\begin{aligned}
f(x,\theta) =\;& a + \sum_{k=1}^{H_2} b_k \,
\tanh \Bigg(
    c_k + \sum_{j=1}^{H_1} w_{kj} \\
& \times \tanh \Bigg(
    d_j +\sum_{i=1}^{l} v_{ji} x_i
\Bigg)
\Bigg),
\end{aligned}
\endgroup
\end{equation}
where \( x_i = (x_1, x_2, \ldots, x_l) \) represents the input vector of \( l \) features, initially limited to the proton number (\( Z \)), the mass number (\( A \)) of the fission fragments and the excitation energy (\( E \)) of the compound nucleus in this work. In this work, the excitation energy is defined as the sum of the incident neutron energy and the neutron separation energy of the target nucleus. $\theta = \{a, b_k, c_k, d_j, w_{kj}, v_{ji}\}$ denotes the set of trainable parameters in the BNN. Specifically, $v_{ji}$ and $w_{kj}$ are the connection weights between the input layer and the first hidden layer, and between the first and second hidden layers, respectively. $d_j$ and $c_k$ are the corresponding bias terms of the first and second hidden layers. $b_k$ denotes the connection weight between the second hidden layer and the output layer, while $a$ is the bias term of the output layer. 

For a new input \(x_n\), the predictive result is obtained by marginalizing over the posterior distribution of the model parameters. The Bayesian predictive distribution is given by,
\begin{equation}
\mathrm{p}(t_n\mid x_n,D)
=
\int
\mathrm{p}(t_n\mid x_n,\theta)
\mathrm{p}(\theta\mid D)
\, \textrm{d}\theta.
\end{equation}
Each prediction corresponding to a specific set of parameters is weighted by its posterior probability, and the integration yields the final predictive output of the BNN.

In practical calculations, the analytical solution of the posterior distribution in Bayesian neural networks is generally intractable due to the high dimensionality of the parameter space. Therefore, the Markov Chain Monte Carlo (MCMC) method is employed in this work to infer the posterior distribution of the network parameters. This approach constructs a Markov chain whose stationary distribution corresponds to the target posterior distribution. After sufficient iterations and convergence of the Markov chain, posterior samples of the model parameters are obtained and used to generate predictive distributions for fission yields \cite{neal2012mcmc}. For each input sample, the predictive mean is calculated by averaging the outputs over the posterior parameter samples, while the predictive uncertainty is quantified from the corresponding posterior predictive distribution. The 95\% confidence interval (CI) is defined by the 2.5th and 97.5th percentiles of the predictive distribution. In this way, the high-dimensional integrals involved in Bayesian inference can be approximated by sample averages, enabling efficient uncertainty quantification and prediction. To quantitatively evaluate the reliability of the predictive uncertainty, we compute the empirical coverage of the confidence intervals. The empirical coverage is defined as
\begin{equation}
\text{CI Coverage} = \frac{N_{\text{inside}}}{N_{\text{total}}},
\end{equation}
where \(N_{\text{inside}}\) denotes the number of samples whose target values fall within the predicted 95\% confidence intervals, and \(N_{\text{total}}\) is the total number of data points in the training dataset, including both evaluated and experimental fission-yield data. This metric provides a direct assessment of the reliability of the predicted confidence intervals.

Given \(S\) posterior samples, the predictive mean can be approximated by the Monte Carlo average,
\begin{equation}
\begingroup
\begin{aligned}
\langle f(x_n) \rangle
\approx
\frac{1}{S}
\sum_{s=1}^{S}
f(x_n,\theta^{(s)}),
\end{aligned}
\endgroup
\end{equation}
for the S groups of posterior predictive samples obtained from the model prediction, the lower and upper bounds of the 95\% confidence interval of the Bayesian neural network can be determined from the 2.5th and 97.5th percentiles of the predictive distribution, respectively.

A conventional random split of individual data points was not adopted in this work because fission-yield data at the same incident neutron energy are strongly correlated. Such a split may place closely related samples into both training and testing subsets, leading to an overestimation of the model generalization capability. For $^{235}$U, all available data were used for training to preserve the completeness of the isotopic-yield distributions. For $^{238}$U, the dataset at 4.6 MeV was completely excluded from training and used as an independent validation dataset, while all remaining data were included in the training set. The predictive capability of the model was further evaluated through the energy dependence of representative cumulative fission yields at neutron energies not explicitly used during training.

The present work aims to improve the model performance through two main strategies: the incorporation of additional physical features and refined data preprocessing. Initially, the \( x_i \) consisted of three inputs: the proton number (\( Z_{\text{in}} \)) and mass number (\( A_{\text{in}} \)) of the fission fragments, and the excitation energy (\( E_{\text{in}} \)) of the compound nucleus. 
The network trained on this initial input is referred to as the BNN$_0$ model.

Subsequently, additional physical features were progressively incorporated into the training dataset, including the odd-even effect, the $\beta$-decay energy, and the isospin. These physical quantities were introduced as additional input features to the BNN training dataset in order to provide supplementary nuclear-structure information relevant to fission yield systematics. In the fission process, nucleon pairing correlations play an important role in determining the stability and yield distribution of fission fragments. Since like nucleons tend to form paired configurations, breaking a proton pair or a neutron pair requires additional pairing energy.
 
To quantify the individual contributions of the physics-informed features, an ablation study was performed by progressively introducing the odd-even effect, isospin, and $\beta$-decay energy into the baseline BNN model. The results reported in Ref.~\cite{wyx} demonstrate that the incorporation of physics-informed features, particularly the odd-even effect and isospin, significantly improves the descriptive capability and predictive accuracy of the BNN model for fission-yield distributions. Specifically, these features enable a more accurate reproduction of the fine structures around the peaks of the double-humped independent mass-yield distributions. In addition, the inclusion of these physical features improves the model's ability to capture the energy dependence of fission yields, leading to a more consistent description of mass-yield variations with incident neutron energy.
\begin{table}[htbp]
\centering
\caption{Comparison of the prediction performance of different Bayesian neural network models. BNN$_0$-IND and BNN$_0$-CUM denote the baseline models for independent and cumulative fission yields, respectively. BNN$_{1-1}$ includes the odd-even effect, BNN$_{1-2}$ includes the isospin, and BNN$_2$ incorporates the odd-even effect and isospin.}
\label{tab:information}
\begin{tabular}{lcc}
\toprule
Model & Yield type & MSE \\
\midrule
BNN$_0$-IND & IND & 0.0491 \\
BNN$_{1-1}$   & IND & 0.0399 \\
BNN$_{1-2}$   & IND & 0.0717 \\
BNN$_2$-IND & IND & 0.0315 \\
\midrule
BNN$_0$-CUM & CUM & 0.610 \\
BNN$_2$-CUM & CUM & 0.441 \\
\bottomrule
\end{tabular}
\end{table}
Table \ref{tab:information} presents the ablation study of the proposed physics-informed features. The inclusion of the odd-even effect alone reduces the MSE from 0.0491 to 0.0399, corresponding to an improvement of 18.7\%, indicating that pairing-related structural information plays an important role in describing independent fission yields. In contrast, the model using only the isospin feature exhibits a larger MSE (0.0717), suggesting that neutron richness alone is insufficient to characterize the complex behavior of fission-product distributions. When the odd-even effect and isospin are incorporated simultaneously, the MSE further decreases to 0.0315, representing a total reduction of 35.8\% relative to the baseline model. This result indicates that the two physics-informed features provide complementary information: the odd-even effect reflects pairing correlations, while the isospin parameter characterizes neutron richness. Their combination enables a more comprehensive description of isotopic fission-yield systematics. For cumulative fission yields, the proposed physics-informed model reduces the MSE from 0.610 to 0.441, corresponding to an improvement of 27.7\%. This demonstrates that the introduced physical features enhance not only the prediction of independent yields but also the reconstruction of cumulative yields through radioactive decay chains.

The treatment of the odd-even effect, denoted as $\delta_{np}$, follows the approach proposed in Ref.~\cite{wangn_com_2022}, and is given by 
\begin{equation}
\delta_{np} = 
\begin{cases}
2 - |I|, & \text{if both } N \text{ and } Z \text{ are even} \\
|I|, & \text{if both } N \text{ and } Z \text{ are odd} \\
1 - |I|, & \text{if } N \text{ is even, } Z \text{ is odd, and } N > Z \\
1 - |I|, & \text{if } N \text{ is odd, } Z \text{ is even, and } N < Z \\
1, & \text{if } N \text{ is even, } Z \text{ is odd, and } N < Z \\
1, & \text{if } N \text{ is odd, } Z \text{ is even, and } N > Z \\
\end{cases},
\end{equation}
where $I$ is defined as $I=(N-Z)/A$. During the strong-interaction dominated stage of nuclear fission, the third component of isospin, $T_z=(N-Z)/2$, remains conserved in the absence of proton-neutron conversion processes. To incorporate isospin-related information into the BNN model, an additional input parameter, $T_3$, was introduced and defined as $(Z - N)/50$.

Fission fragments produced immediately after scission are typically in highly excited and neutron-rich states, and subsequently undergo prompt neutron emission, $\gamma$-deexcitation, and radioactive decay processes. Cumulative fission yields include not only the direct production contribution of a given nuclide, but also the contributions fed by the decay chains of its precursor nuclides. Since fission fragments are generally far from the valley of stability, $\beta$-decay usually dominates the subsequent decay processes. 

To investigate the influence of the $\beta$-decay energy on the predictive performance of the network, the $\beta$-decay energy was independently introduced into the original training dataset as an additional input feature, and the BNN$_\beta$ model was constructed. The $\beta$-decay energy was incorporated through the parameter $Q_{\beta,\text{in}}$ (MeV), defined as $({Q_\beta - 5.5})/{15}$. The $\beta$-decay energy ($Q_\beta$) used in this work was obtained from the National Nuclear Data Center (NNDC) database. All nuclides included in the training and prediction datasets have corresponding $Q_\beta$ values available. Therefore, no missing-value treatment, interpolation, or imputation procedure was required during dataset construction. The absolute deviations between the BNN predictions and the evaluated data are shown in Fig.~\ref{beta1}. The results indicate that the incorporation of the $\beta$-decay energy significantly improves the descriptive capability of the BNN model for fission yield distributions. The total error decreases from 0.00474 to 0.00287, corresponding to a reduction of 39.64\%. The discrepancies between the predicted results and the target data are substantially reduced, leading to a remarkable decrease in the mean squared error (MSE), defined as 
\begin{equation} 
\mathrm{MSE}=\frac{1}{N}\sum_{i=1}^{N}\left(\langle f(x_n) \rangle-t_i\right)^2, 
\end{equation} 
where \(\langle f(x_n) \rangle\) and \(t_i\) denote the predicted and target fission yields, respectively. Unless otherwise specified, all errors discussed hereafter are calculated using this MSE metric in linear yield space. The uncertainties of the target data are incorporated through the Bayesian likelihood function and therefore naturally contribute to the parameter inference process. 
Based on the above analyses, a BNN$_3$ model incorporating the odd-even effect, isospin, and $\beta$-decay energy simultaneously as additional physical input features was constructed. In the following sections, the evaluation results of the BNN$_3$ model for neutron-induced cumulative fission yields of $^{235,238}$U are presented and discussed in detail.

\section{Results and discussion}\label{results}

\subsection{BNN Evaluation Results of Fission Yields for $^{235}$U}\label{results-U235}
The training dataset was constructed from both experimental fission-yield data and evaluated nuclear-data libraries. Experimental data provide direct measurements of fission yields, while evaluated libraries provide more complete and internally consistent isotopic-yield distributions over a wider range of nuclides. The combination of these two data sources improves both the coverage and the robustness of the training dataset.

No explicit source-dependent weighting factors were introduced. Instead, all data points contribute through the Bayesian likelihood function according to their associated uncertainties. Consequently, data with smaller uncertainties impose stronger constraints on the posterior parameter distributions. Owing to the limited amount of available fission-yield data and the energy-based validation strategy adopted in this work, separate training, validation, and test MSE values are not reported. Instead, model performance is assessed through the reconstruction of isotopic-yield distributions, the independent 4.6 MeV validation dataset, and the energy-dependent yield predictions.

For the evaluation of $^{235}$U fission product yields using the BNN model, the training dataset consisted of 3,096 data points from the JENDL evaluated library \cite{jendl-SHIBATA01012011} at thermal, 0.5 MeV, and 14 MeV incident neutron energies, and 1,521 experimental data points extracted from the EXFOR database \cite{exfor}. The BNN$_0$ and BNN$_3$ models were each trained for $10^5$ iterations.

Figure~\ref{fig2-U235-heatmap} illustrates the cumulative fission product yields ($Y_{\mathrm{CUM}}$) for $^{235}\mathrm{U}$ at three representative incident neutron energies, comparing the JENDL evaluated data ($Y_{\mathrm{JENDL}}$) with the predictions of two Bayesian neural networks: the baseline model trained on the original dataset ($Y_{\mathrm{BNN_0}}$) and the physics-informed model incorporating the odd--even effect, isospin, and $\beta$-decay energy ($Y_{\mathrm{BNN_3}}$). In all cases, the characteristic asymmetric double-humped yield distribution is observed, with the light fragment peak around $Z=34$--$43$ and the heavy fragment peak around $Z=51$--$60$. 
Comparing across models, $\mathrm{BNN_0}$ captures the overall double-peak structure but exhibits smoothed local fluctuations and underestimates the odd--even staggering of yields. In contrast, $\mathrm{BNN_3}$ closely reproduces both the global shape and local features of $Y_{\mathrm{JENDL}}$, with sharper peaks and a more pronounced odd--even effect, particularly in the symmetric fission region and near magic numbers. These results indicate that incorporation of physics-informed features significantly enhances the predictive capability of the Bayesian neural network, enabling it to reproduce not only the large-scale fission patterns but also the finer nuclear-structure-dependent yield variations. 
The corresponding residual distributions shown in Fig.~\ref{fig-235-error} further demonstrate the superiority of the physics-informed model. For thermal-neutron, 0.5 MeV, and 14 MeV induced fission, the errors of $\mathrm{BNN_0}$ are characterized by pronounced localized deviations in both the light and heavy fragment regions, especially along isotopic chains exhibiting strong odd--even staggering. After incorporating the odd--even effect, isospin, and $\beta$-decay energy, $\mathrm{BNN_3}$ exhibits substantially reduced error amplitudes and a more uniform error distribution across the nuclide chart. Most large deviations observed in $\mathrm{BNN_0}$ are significantly suppressed, and the residual errors are concentrated near zero. The improvement is particularly evident at thermal and 0.5 MeV neutron energies. Even at 14 MeV, where the fission-fragment distribution becomes broader and more complex, $\mathrm{BNN_3}$ maintains noticeably smaller deviations from the JENDL evaluated data. These results confirm that the introduced physics-informed features enhance both the global predictive accuracy and the local reliability of the Bayesian neural network.

Figures~\ref{fig3-U235-0.0253eV}--\ref{fig5-U235-14MeV} present the isotopic chain yield distributions of fission fragments produced in $^{235}\mathrm{U}$ fission induced by thermal, 0.5~MeV, and 14~MeV neutrons. In Fig.~\ref{fig3-U235-0.0253eV}(b), for $Z=39$, $Y_\mathrm{BNN_0}$ exhibits an overly smooth tail compared to the evaluated data, while $Y_\mathrm{BNN_3}$ closely follows the trend of $Y_\mathrm{JENDL}$, capturing the minor fluctuations more accurately. For $Z=40$, in the neutron number range $N=50$--53, $Y_\mathrm{BNN_0}$ shows oscillatory behavior not present in the evaluated data, and between $N=58$--60, it significantly underestimates the yields relative to $Y_\mathrm{JENDL}$. For $Z=41$, the BNN$_0$ predictions in the oscillatory region deviate markedly from $Y_\mathrm{JENDL}$, and the trend over $N=56$--64 is overly smooth, whereas BNN$_3$ reproduces the evaluated pattern with high fidelity. At $Z=43$, BNN$_3$ notably improves upon BNN$_0$, correcting the pronounced deviations in the $N=56$--59 region and aligning well with $Y_\mathrm{JENDL}$. For $Z=58$, in the region $N=82$--85, $Y_\mathrm{JENDL}$ exhibits a relatively flat trend, while BNN$_0$ predicts an overall increase, whereas BNN$_3$ closely matches the evaluated distribution. Across other isotopic chains, BNN$_3$ consistently provides enhanced predictive accuracy compared to BNN$_0$, with chain yield distributions closely reproducing the evaluated data. Overall, the results indicate that the incorporation of physics-informed features in BNN$_3$ significantly improves the reproduction of local structures and oscillatory behavior in the isotopic chain yields.

In Fig.~\ref{fig4-U235-0.5MeV}, for $Z=38$, both $Y_\mathrm{BNN_0}$ and $Y_\mathrm{BNN_3}$ exhibit noticeable deviations from the evaluated data in the overall distribution trend; however, $\mathrm{BNN_3}$ provides isotopic yields that are closer in magnitude to those of $Y_\mathrm{JENDL}$ for several nuclides. For $Z=40$, $\mathrm{BNN_0}$ fails to correctly predict the nuclide corresponding to the peak of the chain-yield distribution. In contrast, $Y_\mathrm{BNN_3}$ demonstrates clear improvements over $Y_\mathrm{BNN_0}$ in both the overall distribution trend and reconstruction accuracy, showing notable consistency with $Y_\mathrm{JENDL}$. For $Z=41$, the $Y_\mathrm{JENDL}$ displays a complex oscillatory behavior along the isotopic chain. Although $\mathrm{BNN_0}$ reproduces some of these oscillations, its predictions differ substantially from $Y_\mathrm{JENDL}$ and become overly smooth in the latter part of the distribution. By comparison, $Y_\mathrm{BNN_3}$ successfully captures the oscillatory pattern and exhibits better agreement with $Y_\mathrm{JENDL}$. For $Z=43$, $\mathrm{BNN_3}$ accurately reproduces the oscillatory behavior at $N=56$--59, as well as the corresponding yield magnitudes. For $Z=58$, the evaluated isotopic chain yields show an overall decreasing trend with increasing neutron number. BNN$_0$ instead predicts an initial increase followed by a decrease, whereas $Y_\mathrm{BNN_3}$ correctly captures the overall trend and remains consistent with $Y_\mathrm{JENDL}$. For $Z=59$, $Y_\mathrm{BNN_0}$ significantly overestimates the yield at $N=83$ compared with the evaluated data. Overall, $Y_\mathrm{BNN_3}$ exhibits a chain-yield distribution that is more consistent with the evaluated data than $Y_\mathrm{BNN_0}$ and achieves higher reconstruction accuracy across the isotopic chain.

In Fig.~\ref{fig5-U235-14MeV}, for $Z=38$, $Y_\mathrm{BNN_3}$ shows better overall agreement with $Y_\mathrm{JENDL}$ in the isotopic chain yield distribution, although it slightly overestimates the yield at $N=58$. For $Z=39$, the isotopic chain yields increase and decrease slowly at first and then drop rapidly, with a local maximum appearing at $N=60$. $\mathrm{BNN_3}$ successfully reproduces this overall trend. For $Z=40$, the isotopic chain yields vary relatively smoothly. $Y_\mathrm{BNN_0}$ exhibits excessive oscillations in the predicted distribution, whereas $Y_\mathrm{BNN_3}$ provides a smoother distribution that is more consistent with $Y_\mathrm{JENDL}$. For $Z=41$, $Y_\mathrm{BNN_3}$ successfully reproduces the oscillatory behavior observed in the $Y_\mathrm{JENDL}$ evaluation and achieves higher reconstruction accuracy than $Y_\mathrm{BNN_0}$. For $Z=43$, a local minimum appears in the isotopic chain yield distribution at $N=57$, where $Y_\mathrm{BNN_0}$ overestimates the yield relative to $Y_\mathrm{JENDL}$. For $Z=52$, $\mathrm{BNN_0}$ significantly underestimates the yields in the range of $N=73$--78. At $N=79$, the evaluated data exhibit a local minimum, while $Y_\mathrm{BNN_0}$ predicts an opposite trend.  For $Z=53$, $Y_\mathrm{BNN_3}$ successfully captures the oscillatory behavior at the beginning of the isotopic chain yield distribution. Overall, for the isotopic chain yield distributions of $^{235}\mathrm{U}$ fission induced by 14~MeV neutrons, $\mathrm{BNN_3}$ provides chain yields that are more consistent with the evaluated data and demonstrates an overall improvement in reconstruction accuracy compared with $Y_\mathrm{BNN_0}$.

The variation of the fission product yields with incident neutron energy reflects the combined effects of energy partitioning, mass split modes, and subsequent decay processes in the fission mechanism. This energy dependence is a fundamental issue in fission yield studies and is of practical importance in reactor design, nuclear data evaluation, and radioactive waste management. Figure~\ref{fig6-U235-yieldenergy} presents the energy dependence of 16 representative nuclides produced in neutron-induced fission of $^{235}$U over the energy range of 0-20~MeV. The performance of two BNN models is compared with experimental data ($Y_\mathrm{EXP}$) and evaluated data from the JENDL library ($Y_\mathrm{JENDL}$). As shown in Fig.~\ref{fig6-U235-yieldenergy}(b), the experimental yields exhibit a continuously decreasing trend with increasing incident neutron energy, characterized by a rapid decrease at low energies followed by a more gradual reduction. ${\mathrm{BNN}_0}$ predicts an almost constant-slope behavior and systematically overestimates the experimental yields. Although ${\mathrm{BNN}_3}$ slightly overpredicts the yields in the 0--9~MeV range, it reproduces the overall experimental distribution more closely than $Y_{\mathrm{BNN}_0}$. In Fig.~\ref{fig6-U235-yieldenergy}(c), the yields decrease monotonically with energy. Both $Y_{\mathrm{BNN}_0}$ and $Y_{\mathrm{BNN}_3}$ follow trends consistent with the evaluated data; however, in the intermediate energy range of 3--9~MeV, $Y_{\mathrm{BNN}_3}$ shows better agreement with the experimental data. In Figs.~\ref{fig6-U235-yieldenergy}(c--e), at incident neutron energies of 14--15~MeV, $Y_{\mathrm{BNN}_3}$ remains consistent with both experimental and evaluated data. For Fig.~\ref{fig6-U235-yieldenergy}(g), corresponding to $^{140}$Ba, both the experimental and evaluated data are largely encompassed within the confidence intervals predicted by the BNN models. $Y_{\mathrm{BNN}_0}$ exhibits a relatively smooth variation with energy, whereas $Y_{\mathrm{BNN}_3}$ shows a steeper decreasing trend, with a rapid reduction in yields in the 14--20~MeV energy range.  In Fig.~\ref{fig6-U235-yieldenergy}(h), $Y_{\mathrm{BNN}_3}$ agrees well with $Y_{\mathrm{EXP}}$ in the 0--6~MeV region, while ${\mathrm{BNN}_0}$ significantly underestimates the yields relative to $Y_{\mathrm{JENDL}}$. In Fig.~\ref{fig6-U235-yieldenergy}(j), $Y_{\mathrm{BNN}_0}$ shows an increasing trend followed by a slight decrease, whereas $Y_{\mathrm{BNN}_3}$ increases monotonically and exceeds both $Y_{\mathrm{JENDL}}$ and $Y_{\mathrm{EXP}}$ at 14--15~MeV. In Fig.~\ref{fig6-U235-yieldenergy}(k), $Y_{\mathrm{BNN}_0}$ increases and then remains nearly constant, while $Y_{\mathrm{BNN}_3}$ exhibits a continuously increasing trend, with both $Y_{\mathrm{JENDL}}$ and $Y_{\mathrm{EXP}}$ well covered by the predicted confidence intervals. In Fig.~\ref{fig6-U235-yieldenergy}(m), the experimental yields show an increase followed by a decrease with energy. $Y_{\mathrm{BNN}_0}$ decreases monotonically, whereas $Y_{\mathrm{BNN}_3}$ reproduces the rising-and-falling behavior and agrees well with both $Y_{\mathrm{JENDL}}$ and $Y_{\mathrm{EXP}}$ at 14--15~MeV, although it slightly underestimates the experimental data in the 0--8~MeV range. In Fig.~\ref{fig6-U235-yieldenergy}(n), $Y_{\mathrm{BNN}_3}$ more closely follows the increasing trend observed in the experimental data. For all nuclides shown, the confidence intervals predicted by BNN$_3$ are consistently narrower than those of BNN$_0$. The incorporation of additional physical constraints in BNN$_3$ leads to improved predictive accuracy and a significant reduction in predictive uncertainty, demonstrating that the inclusion of physical information effectively enhances the model performance.

\subsection{BNN Evaluation Results of Fission Yields for $^{238}$U}\label{results-U238}

For the evaluation of $^{238}$U fission product yields using the BNN model, the training dataset consisted of 2064 data points from the JENDL evaluated library \cite{jendl-SHIBATA01012011} of 0.5 and 14 MeV neutrons, and 1534 experimental data points extracted from the EXFOR database \cite{exfor}. The BNN$_0$ and BNN$_3$ models were each trained for $10^5$ iterations. Figure~\ref{fig7-U238-heatmap} shows the cumulative fission product yield distributions of the \(^{238}\mathrm{U}(n,f)\) reaction at neutron energies of 0.5~MeV and 14~MeV. The \(Y_{\mathrm{JENDL}}\) exhibit the characteristic double-humped structure of asymmetric fission at low energy, with an increased contribution from the symmetric fission region at 14~MeV. Both BNN models reproduce the overall asymmetric yield pattern and the locations of the light and heavy fragment groups. The BNN$_0$ results, however, show a more diffuse distribution near the yield ridges, particularly in the region \(Z=36\)--46 and \(N=50\)--63, where fine structural features are smeared. In contrast, BNN$_3$ provides sharper peak contours and improved agreement with \(Y_{\mathrm{JENDL}}\). 
Figure~\ref{fig-238-error} presents the residual distributions between the BNN predictions and the JENDL evaluated data, providing additional insight into the predictive performance of the two models. At 0.5~MeV, the BNN$_0$ model exhibits pronounced localized deviations around both the light and heavy fragment yield peaks, where alternating overestimations and underestimations indicate difficulties in reproducing fine yield fluctuations and odd--even staggering effects. In comparison, BNN$_3$ significantly suppresses these deviations, resulting in smaller error amplitudes and a more uniform error distribution over the nuclide chart. At 14~MeV, where the contribution of symmetric fission becomes more important and the yield distribution broadens, BNN$_0$ shows systematic deviations along the main fission-fragment valley. These deviations are substantially reduced in BNN$_3$, which maintains closer agreement with the evaluated data across both asymmetric and symmetric fission regions. The reduced residual magnitude and improved spatial consistency demonstrate that the incorporation of physically informed features enhances the model's ability to capture both local nuclear-structure effects and energy-dependent variations in cumulative fission yields. Overall, the inclusion of physical constraints in BNN$_3$ improves the description of local yield structures and energy-dependent effects, leading to enhanced physical consistency and predictive reliability in fission yield evaluation.

Figure~\ref{fig8-U238-0.5MeV} presents the isotopic yield distributions of \(^{238}\)U fission induced by 0.5~MeV neutrons for charge numbers \(Z=38\)--44 and 52--60. The evaluated data \(Y_{\mathrm{JENDL}}\) exhibit well-defined isotopic systematics, including pronounced peak structures, local minima, and oscillatory behavior associated with nuclear structure effects. While BNN$_0$ reproduces the overall trends of the isotopic chains, it shows systematic deficiencies in resolving fine structures, leading to smoothed peak regions and inaccurate descriptions of local yield variations, particularly in the light and intermediate fragment regions. In contrast, BNN$_3$ consistently restores the detailed shapes of the isotopic distributions, accurately reproducing peak positions, local fluctuations, and rapid yield changes across a broad range of charge numbers. This improvement is especially evident in regions exhibiting strong oscillatory patterns and narrow yield maxima. Overall, the incorporation of physical constraints in BNN$_3$ significantly enhances the model’s ability to capture detailed isotopic yield systematics, resulting in improved predictive performance and increased physical consistency in the evaluation of fission product yields.

Figure~\ref{fig9-U238-14MeV} presents the isotopic yield distributions of \(^{238}\)U fission induced by 14~MeV neutrons across a wide range of charge numbers. BNN$_0$ captures the overall distributions, correctly reproducing the general trend of rising and falling yields across each chain. However, it fails to reproduce the detailed local fluctuations present in the \(Y_{\mathrm{JENDL}}\), producing overly smooth curves that omit physically meaningful variations such as small peaks, valleys, and odd--even oscillations. In contrast, BNN$_3$, which incorporates physical constraints, preserves these local features while maintaining smoothness and continuity along the isotopic chains. This results in a more faithful representation of both global and local characteristics of the fission yield distributions. Overall, the comparison demonstrates that including nuclear physics constraints in the BNN framework enhances the model’s ability to reproduce fine isotopic structures, yielding more reliable and physically consistent predictions. 

\begin{table}[htbp]
    \centering
    \caption{Number of fission fragments with available experimental data for neutron-induced fission of $^{238}$U at different incident neutron energies.}
    \label{table:exp238}
    \begin{tabular}{cccccc}
        \hline
        $e_n$/MeV & counts & $e_n$/MeV & counts & $e_n$/MeV & counts \\
        \hline
        0.0000005 & 48 & 3.72  & 31 & 8.1   & 28  \\
        0.4       & 6  & 3.726 & 29 & 8.53  & 33  \\
        0.5       & 9  & 3.75  & 11 & 8.9   & 7    \\
        0.7       & 25 & 3.9   & 3  & 9.1   & 30   \\
        1         & 50 & 4.2   & 11 & 9.35  & 33   \\
        1.01      & 11 & 4.49  & 14 & 10.09 & 32    \\
        1.3       & 16 & 4.6   & 94 & 11.3  & 40    \\
        1.37      & 12 & 4.7   & 4  & 12.52 & 33   \\
        1.4       & 6  & 4.72  & 11 & 13.4  & 5   \\
        1.5       & 3  & 4.782 & 30 & 14    & 6   \\
        1.52      & 25 & 5     & 35 & 14.1  & 40   \\
        1.722     & 29 & 5.42  & 32 & 14.3  & 24   \\
        1.9       & 90 & 5.5   & 3  & 14.5  & 14    \\
        2         & 3  & 5.982 & 29 & 14.7  & 49    \\
        2.16      & 29 & 6     & 28 & 14.8  & 53      \\
        2.37      & 14 & 6.35  & 33 & 14.9  & 43    \\
        2.86      & 15 & 6.9   & 2  & 17.3  & 4        \\
        3         & 44 & 7.1   & 28 & 18.1  & 2       \\
        3.23      & 11 & 7.7   & 2        \\
        3.6       & 14 & 7.75  & 32   \\
        \hline
    \end{tabular}
\end{table}

Figure~\ref{fig10-U238-4.6MeV} shows the BNN$_0$ and BNN$_3$ predictions of the isotopic yield distributions for \(^{238}\)U fission at 4.6~MeV neutron energy, which was not included in the training set, compared with the experimental data (\(Y_{\mathrm{EXP}}\)). Among all available experimental energy points listed in Table~\ref{table:exp238}, the 4.6~MeV dataset contains the largest number of measured fission fragments, providing the most complete isotopic coverage and the strongest statistical representativeness. Therefore, this energy point offers a particularly stringent benchmark for evaluating the predictive capability and generalization performance of the BNN models outside the training domain. Overall, both BNN$_0$ and BNN$_3$ successfully reproduce the global shape of the isotopic yield chains, including the peak positions and the general width of the distributions for both light and heavy fragments, although certain deviations in the absolute yield values are still observed. In particular, noticeable differences between the two models appear in Fig.~\ref{fig10-U238-4.6MeV}(h) and Fig.~\ref{fig10-U238-4.6MeV}(l), where BNN$_3$ exhibits stronger local oscillatory behavior and larger yield variations, while the predictions of BNN$_0$ remain comparatively smooth. This behavior is consistent with the characteristics of the training data, in which the evaluated yield distributions at the training energy points of 0.5 and 14~MeV exhibit pronounced local fluctuations for the isotopic chains with proton numbers Z=43 and Z=55. Overall, the BNN$_3$ predictions are generally consistent with the trends observed in the available experimental data. Nevertheless, owing to the incomplete isotopic coverage and relatively large uncertainties of the experimental measurements, experimental data alone cannot provide a sufficiently comprehensive reference for assessing isotopic-chain yield distributions at a given incident neutron energy. Evaluated nuclear data libraries, which provide more complete and internally consistent isotopic yield distributions, therefore constitute an important complementary benchmark for model assessment. Under this criterion, BNN$_3$ demonstrates improved consistency with the evaluated isotopic-chain distributions and their energy-dependent evolution, while maintaining reasonable agreement with the available experimental observations.

Figure~\ref{fig11-U238-yieldenergy} presents the neutron energy dependence of cumulative fission product yields for 16 representative nuclides from $^{238}$U(n,f) over the range from 0 to 15~MeV. In Fig.~\ref{fig11-U238-yieldenergy}(c) for $^{99}$Mo, the yield variation predicted by BNN$_3$ is in good agreement with both the evaluated data and the overall distribution of the experimental measurements, whereas $Y_{\mathrm{BNN_0}}$ is clearly lower than $Y_{\mathrm{EXP}}$ over most of the energy range. For $^{103}$Ru shown in Fig.~\ref{fig11-U238-yieldenergy}(g), both BNN$_0$ and BNN$_3$ predict a gradual decrease in yield with increasing incident neutron energy; however, the yield values predicted by BNN$_3$ show better agreement with the experimental data in the 0--5~MeV energy range. In the cases of $^{141,143,144}$Ce presented in Figs.~\ref{fig11-U238-yieldenergy}(m)--(o), both the experimental and evaluated data are largely encompassed by the confidence intervals predicted by the BNN models. The baseline model BNN$_0$ exhibits a relatively flat energy dependence, while BNN$_3$ predicts a more pronounced decreasing trend. In particular, for Figs.~\ref{fig11-U238-yieldenergy}(m) and (o), the BNN$_3$ predictions in the 0--4~MeV energy range show improved agreement with the experimental data. Moreover, the 95\% confidence intervals associated with BNN$_3$ are generally narrower than those of BNN$_0$, indicating reduced model uncertainty and improved predictive precision. Overall, despite the intrinsic inconsistencies and limited accuracy of the available experimental data for $^{238}$U, the BNN$_3$ model demonstrates strong generalizability and provides a more reliable and robust evaluation of energy-dependent fission product yields over a wide range of fission products and neutron energies.

\begin{table}[htbp]
    \centering
    \caption{Comparison of the global reconstruction mean squared error (MSE), mean prediction interval width (MPIW), and 95\% confidence interval coverage (CI Coverage) between BNN$_0$ and BNN$_3$ for $^{235}$U and $^{238}$U.}
    \label{tab:error&ci}
    \begin{tabular}{cccc}\hline
                                    &BNN$_0$        &BNN$_3$    &Reduction    \\  \hline
        MSE ($^{235}$U)             & 0.00367         &0.00184      &49.8\%        \\
        MSE ($^{238}$U)             & 0.00571        & 0.00290     &49.2\%        \\  
        MPIW ($^{235}$U)            &1.739          &1.302     &25.1\%         \\ 
        MPIW ($^{238}$U)            &2.189          &1.626      &25.7\%         \\
        CI Coverage ($^{235}$U)     & 95.50\%       & 97.20\%   &             \\   
        CI Coverage ($^{238}$U)     & 96.22\%       & 95.89\%   &             \\ \hline
    \end{tabular}
\end{table}

Table~\ref{tab:error&ci} presents the comparison of the mean squared error (MSE), mean prediction interval width (MPIW), and 95\% confidence interval coverage (CI Coverage) between BNN$_0$ and BNN$_3$ for $^{235}$U and $^{238}$U. The MSE is used to evaluate the reconstruction accuracy of the model and represents the global reconstruction error between the predicted and target cumulative fission-yield distributions. Calculated over all available data points, it provides a quantitative measure of the overall agreement between the predicted and target yield distributions, with lower MSE values indicating a more accurate reproduction of the systematic features of cumulative fission yields. The MPIW reflects the sharpness of the predictive intervals, and the CI Coverage indicates the proportion of target data covered by the corresponding 95\% confidence intervals. Compared with BNN$_0$, BNN$_3$ achieves substantial reductions in the MSEs for both $^{235}$U and $^{238}$U, with decreases of 49.8\% and 49.2\%, respectively, indicating a significant improvement in reconstruction accuracy. Meanwhile, the MPIW values are reduced by 25.1\% and 25.7\%, respectively, demonstrating that the improved model provides narrower predictive intervals. At the same time, the CI Coverage values remain close to the nominal confidence level, indicating that the improved model maintains reliable uncertainty quantification capability while enhancing predictive performance. These results suggest that BNN$_3$ provides narrower predictive intervals without compromising the reliability of uncertainty quantification.

\begin{table}[htbp]
\centering
\caption{Comparison of the prediction errors between BNN$_0$ and BNN$_3$ for different yield regions of $^{235}$U and $^{238}$U.}
\label{tab:reduction}
\begin{tabular}{ccS[table-format=1.6]S[table-format=1.6]c}
\toprule
Nuclide & Yield region & {BNN$_0$} & {BNN$_3$} & Reduction \\
\midrule
\multirow{3}{*}{$^{235}$U}
& $Y \geq 0.1$& 0.00539& 0.00368& 37.96\% \\
& $0.1 > Y \geq 0.001$& 0.00198& 0.00123& 37.95\% \\
& $Y < 0.001$& 0.00212& 0.000150& 92.89\% \\
\midrule
\multirow{3}{*}{$^{238}$U}
& $Y \geq 0.1$& 0.00777& 0.00517& 33.48\% \\
& $0.1 > Y \geq 0.001$& 0.000835& 0.000470& 43.76\% \\
& $Y < 0.001$& 0.00493& 0.000330& 93.30\% \\
\bottomrule
\end{tabular}
\end{table}

Further insight into the predictive improvement can be obtained from Table~\ref{tab:reduction}, which compares the prediction errors of BNN$_0$ and BNN$_3$ in different fission-yield regions. The yield distributions are divided into three intervals according to the magnitude of the cumulative yields: high-yield fragments ($Y \geq 0.1$), intermediate-yield fragments ($0.1 > Y \geq 0.001$), and low-yield fragments $Y < 0.001$). For both $^{235}$U and $^{238}$U, BNN$_3$ consistently outperforms BNN$_0$ across all yield regions. In the high-yield region, the prediction errors are reduced by approximately 33\%--38\%, indicating that the improved model enhances the reproduction of the dominant fission-product peaks and the overall structure of the mass distributions. In the intermediate-yield region, the reductions remain significant, reaching about 37\%--44\%, and the most remarkable improvement appears in the low-yield region, where the prediction errors decrease by more than 92\% for both $^{235}$U and $^{238}$U. Because the cumulative fission yields in this region are extremely small, relative errors can be strongly amplified by small denominators and may therefore provide a misleading measure of model performance. Moreover, some cumulative yields in the dataset reach values as low as ($10^{-30}$), such that directly introducing a logarithmic offset based on the minimum nonzero yield would excessively emphasize numerically negligible yields. Therefore, the error analysis presented here is based on absolute prediction errors, which provide a more robust and physically meaningful assessment of model performance in the low-yield region. Since low-yield isotopes usually exhibit stronger local fluctuations, larger relative uncertainties, and more sparse experimental constraints, accurate prediction in this region is considerably more challenging. The substantial reduction of errors therefore indicates that BNN$_3$ possesses a significantly enhanced capability to learn weak-yield structures and complex local variations from the evaluated data. Combined with the results in Table~\ref{tab:error&ci}, these observations demonstrate that BNN$_3$ not only improves the global reconstruction accuracy and uncertainty sharpness, but also achieves particularly strong performance in describing low-yield fission fragments, which are often the most difficult part of fission-yield evaluations.

\section{Summary}\label{summary}

In this study, an improved Bayesian neural network model (BNN$_3$) was developed based on a baseline architecture consisting of two hidden layers with 20 neurons each (BNN$_0$). The improvements were achieved by incorporating additional physics-informed features, including the odd--even effect, $\beta$-decay energy, and isospin, into the training process, thereby enhancing the physical constraints on the fission yield predictions. A comparative analysis between $Y_{\mathrm{BNN}_0}$ and $Y_{\mathrm{BNN}_3}$ demonstrates that BNN$_3$ exhibits clear improvements in both predictive accuracy and uncertainty quantification, and more reliably reproduces the global distribution characteristics of fission yields as well as the structural patterns along isotopic chains. To further examine the predictive behavior of the model beyond the training dataset, the neutron-induced fission of $^{238}$U at 4.6 MeV, an energy point with relatively comprehensive experimental coverage, was reserved as an independent validation case. The results show that the BNN$_3$ predictions remain broadly consistent with the available experimental observations and reproduce the local structures of the isotopic yield distributions more accurately than BNN$_0$. Although the overall validation error is not reduced at this energy, BNN$_3$ exhibits improved agreement with the corresponding evaluated isotopic yield distributions, suggesting that the incorporated physics-informed features enhance the model’s ability to capture local nuclear-structure effects and systematic yield trends. Further analysis of the energy dependence of fission product yields indicates that BNN$_3$ is able to capture the overall energy-dependent trends while providing significantly reduced uncertainty intervals. These findings demonstrate that the integration of physics-informed features with Bayesian machine learning constitutes an effective approach for improving the reliability of fission yield predictions and can provide valuable support for high-precision nuclear data evaluation.

\begin{acknowledgements}
This work was supported by the National Natural Science Foundation of China (Nos. 12247126 and Nos. 12375123), Henan Postdoctoral Foundation (No. HN2024013), and the Natural Science Foundation of Henan Province (No. 242300421048).
\end{acknowledgements}
\bibliography{fop}
\bibliographystyle{fop}

\end{document}